\documentclass[preprint]{aastex}

\usepackage{color}
\usepackage{rotating}
\usepackage{lscape}
\usepackage{url} % Added by YI
\usepackage{ulem} % Added by YI

\shorttitle{Gamma-ray emitting radio galaxies}
\shortauthors{Y. Fukazawa et al.}

\begin{document}

\title{High energy emission component, population, and contribution to the extragalactic gamma-ray background of gamma-ray emitting radio galaxies}

\author{Yasushi Fukazawa\altaffilmark{1,2,3}, 
Hiroto Matake\altaffilmark{1},
Taishu Kayanoki\altaffilmark{1},
Yoshiyuki Inoue\altaffilmark{4, 5, 6},
Justin Finke\altaffilmark{7}
}

\email{\texttt{fukazawa@astro.hiroshima-u.ac.jp}}

\altaffiltext{1}{Department of Physical Science, Hiroshima University, 1-3-1 Kagamiyama, Higashi-Hiroshima, Hiroshima 739-8526, Japan}
\altaffiltext{2}{Hiroshima Astrophysical Science Center, Hiroshima University, 1-3-1 Kagamiyama, Higashi-Hiroshima, Hiroshima 739-8526, Japan}
\altaffiltext{3}{Core Research for Energetic Universe (Core-U), Hiroshima University, 1-3-1 Kagamiyama, Higashi-Hiroshima, Hiroshima 739-8526, Japan}
\altaffiltext{4}{Department of Earth and Space Science, Graduate School of Science, Osaka University, Toyonaka, Osaka 560-0043, Japan}
\altaffiltext{5}{Interdisciplinary Theoretical \& Mathematical Science Program (iTHEMS), RIKEN, 2-1 Hirosawa, Saitama 351-0198, Japan}
\altaffiltext{6}{Kavli Institute for the Physics and Mathematics of the Universe (WPI), The University of Tokyo, Kashiwa 277-8583, Japan}
\altaffiltext{7}{U.S.\ Naval Research Laboratory, Code 7653, 4555 Overlook Ave.\ SW, Washington, DC, 20375-5352, USA}

\begin{abstract}

In this study, we systematically studied the X-ray to GeV gamma-ray spectra of 61 {\it Fermi} Large Area Telescope (LAT) detected radio galaxies. We found an anticorrelation between peak frequency and peak luminosity in the high-energy spectral component of radio galaxies, similar to blazars. With this sample, we also constructed a gamma-ray luminosity function (GLF) of gamma-ray-loud radio galaxies. We found that blazar-like GLF shapes can reproduce their redshift and luminosity distribution, but the log$N$-log$S$ relation prefers models with more low-$z$ radio galaxies. This indicates many low-$z$ gamma-ray-loud radio galaxies. By utilizing our latest GLF, the contribution of radio galaxies to the extragalactic gamma-ray background is found to be 1--10\%. We further investigated the nature of gamma-ray-loud radio galaxies. Compared to radio or X-ray flux-limited radio galaxy samples, the gamma-ray selected sample tends to lack high radio power galaxies like FR-II radio galaxies.  We also found that only $\sim$10\% of radio galaxies are GeV gamma-ray loud. Radio galaxies may contribute to the cosmic MeV gamma-ray background comparable to blazars if gamma-ray-quiet radio galaxies have X-ray to gamma-ray spectra like Cen A, with a small gamma-ray to X-ray flux ratio.

\end{abstract}
\keywords{galaxies: active ---  galaxies: jets --- X-rays: galaxies}

\section{Introduction}

The formation of supermassive black holes (SMBHs) is one of the most intriguing questions in modern astrophysics. They are believed to coevolve with their host galaxies through feedback processes \citep{Fabian2012}. Active galactic nucleus (AGN) activity drives feedback via relativistic jets or fast outflows. The evolutionary history of these AGN activities is the key to understanding the feedback history in the universe.
The luminosity function (LF) of AGNs and their contribution to the cosmic X-ray and gamma-ray background radiation allow us to unveil the evolutionary history of AGNs. These two are complementary to each other. The former discloses the differential history, while the latter tells the integrated history. 

The evolutionary history of blazars (AGN with jets pointed toward the Earth) and their contribution to the cosmic gamma-ray background (CGB) have been well studied using data from the {\it Neil Gehrels Swift Observatory}'s Burst Alert Telescope (BAT) in hard X-rays and {\it Fermi}/LAT in gamma-rays \citep[see e.g.,][]{Ajel09,Ajel12,Ajel14,Toda20}. However, although blazars are the brightest and most easily detected AGNs with jets due to beaming, they represent $<1$\% of the AGN jet population.  The most numerous AGN jet population is misaligned AGNs (MAGNs), i.e., radio galaxies. No definition of MAGNs are given, but typically MAGNs are sources with jets with viewing angle $>\frac{1}{\Gamma}$, where $\Gamma$ is a jet's Lorentz factor.

Studies of the evolutionary history of radio galaxies have been mainly conducted in the radio band \citep[see, e.g.,][]{Will01} where they are easiest to detect. However, due to synchrotron self-absorption, the radio emission from the relativistic jet is dominated by the downstream component \citep{Blandford1979}, reflecting the past activity of AGNs. High spatial resolution radio observations or gamma-ray observations can allow us to investigate the current jet activity. 

{\it Fermi}/LAT has been monitoring the entire GeV gamma-ray sky since its launch in 2008. Although radio galaxies are several orders of magnitude fainter than blazars in the gamma-ray band, early  {\it Fermi}/LAT observations reported a small number of radio galaxies \citep{1MAGN}. Based on those samples, past studies reported that radio galaxies contribute 10--50\% of the CGB using radio luminosity functions \citep[e.g.,][]{Inou11, DiMa14}.  \citet{Ajel15} argued that together, blazars and star-forming galaxies make up 100\% of the CGB in the 0.1--800~GeV band. However, there is considerable uncertainty in the radio galaxies' contribution due to a limited sample size. An accurate estimate will allow improved estimates of a possible component of the CGB from dark matter annihilation.

Here, the 4th Fermi/LAT Gamma-ray source catalog \citep[4FGL-DR2,][]{Ball20} contains 61 MAGNs based on its 10-yr survey. This number is significantly increased from the past gamma-ray catalog; 11 MAGNs in the first catalog \citep[1FGL,][]{1FGL}, 12 in the second \citep[2FGL,][]{2FGL}, and 21 in the third \citep[3FGL,][]{3FGL}\footnote{Objects classified as {\tt radio galaxy}, {\tt CSS}, and {\tt SSRQ} are included. Six galaxies classified as {\tt other AGN} (almost the same as {\tt AGN} in 4FGL) are listed in the 3FGL}. The larger statistical sample of gamma-ray-detected radio galaxies in 4FGL-DR2 allows us to constrain their gamma-ray LF and their CGB contribution using only gamma-ray data. Others have tried to estimate the radio galaxy contribution to the CGB using radio luminosity functions and ratios of gamma-ray to radio flux \citep[e.g.,][]{Inou11,DiMa14}.

We can do further studies of other features of AGN jets with this statistical sample, such as the relationship between blazars and radio galaxies, and that between gamma-ray selected and radio selected radio galaxies. In the radio band, the radio galaxies are often classified as \citet[][]{Fana74} Type I (FR-I) and Type II (FR-II), where FR-Is are brighter in the middle and fainter at the edges, and FR-IIs are fainter in the middle and brighter at the edges.  \citet{Fana74} and subsequent work \citep[e.g.,][]{Ledl96} indicated that FR-Is are overall fainter in radio than FR-IIs \citep[although see][]{Ming19}.  FR-I radio galaxies
are widely considered the parent population of BL Lac type blazars \citep{Pado90} and FR-II radio
galaxies are considered the parent population of flat spectrum radio quasar (FSRQ) type blazars.  FSRQs are blazars with high luminosities and strong broad emission lines in their optical spectra; while BL Lacs are blazars with lower luminosities and weak or absent broad emission lines.
Recently, radio galaxies with core-dominant radio emission have been found.  These cannot be classified as FR-I or FR-II and have been dubbed FR-0 \citep{Bald10,Ghis11}.
The first {\it Fermi} Gamma-ray Catalog contained 7 FR-Is and 4 FR-IIs \citep{1MAGN}, and thus FR-IIs did not seem to be as bright in the GeV gamma-ray band.
The MAGN sample of the 4FGL-DR2 can elucidate these questions by constraining the gamma-ray LF and comparing it to the gamma-ray LF of blazars, and to the radio LF of radio galaxies.

In addition, we can combine gamma-ray data from {\it Fermi}/LAT with X-ray observations to investigate the spectral energy distribution (SED) shape of radio galaxies from X-rays to gamma-rays. In the case of blazars, the SEDs have a controversial luminosity dependence, the so-called blazar sequence \citep{Fossati1998, Kubo1998, Ghis98, Fink13, Ghisellini2017}, while the SEDs of radio galaxies have not been well studied. Constraining the SED from X-rays to gamma-rays is also essential for estimating the contribution of radio galaxies to the CGB spectrum.

We construct the GLF and the X-ray and gamma-ray SED of radio galaxies and establish their contribution to the CGB. In \S~\ref{sec:sample}  we summarize general properties of our sample of gamma-ray emitting radio galaxies, including their X-ray properties. X-ray and gamma-ray SEDs are presented in \S~\ref{ana:sed}. The gamma-ray luminosity function and the contribution to the CGB is described in \S~\ref{sec:glf} and \S~\ref{sec:cgb}, respectively. The relation to blazars and radio-selected radio galaxies are discussed in \S~\ref{sec:dis}.  Conclusions are presented in \S~\ref{sec:con}. Throughout this paper, we assume cosmological parameters of $\Omega_{\Lambda}=0.7$,  $\Omega_{M}=0.3$ and $H_0$=70 km s$^{-1}$ Mpc$^{-1}$.

\section{Sample of Radio Galaxies}
\label{sec:sample}

\subsection{Gamma-ray data}
{\it Fermi}/LAT has been surveying the gamma-ray sky since 2008 \citep{Atwood2009}. The latest catalog, the 4FGL-DR2, is based on its 10 years of survey data, from 2008 August 4, to 2018 August 2. The previous catalog, the {\it Fermi}/LAT 4th catalog (4FGL), was based on 8 years of data, and was described in detail in \citet{4FGL}. The 4FGL-DR2 has 5788 gamma-ray sources detected at $>4\sigma$ in the energy range from 50~MeV to 1~TeV, including 61 MAGNs. We quote the GeV gamma-ray photon index, gamma-ray flux in 0.1--100 GeV, 6-band flux, redshift, and classification from the 4FGL-DR2. Note that the detection threshold in 4FGL-DR2 is the same as that in the previous catalogs from 1FGL to 4FGL.

The 61 4FGL-DR2 MAGNs are classified into four types; 43 {\tt RDG} (radio galaxies), 5 {\tt CSS} (compact steep spectrum radio sources), 2 {\tt SSRQ} (steep-spectrum radio quasars), and 11 {\tt AGN}, as summarized in Table \ref{sample1}. Hereafter, we call this sample the 4FGL-DR2 MAGNs. Two new radio galaxies ({\tt RDG}) were added in the 4FGL-DR2 that were not in the 4FGL. CSSs are powerful radio sources associated with an AGN, characterized by their small radio size and steep slope of the radio spectrum, with a peak around 100 MHz. SSRQs are high-luminosity AGNs with lobe-dominated radio emission and radio spectral slopes $>$ 0.5 at frequencies of several GHz. {\tt AGN}s as defined in the 4FGL-DR2 are non-blazar AGNs whose existing data do not allow an unambiguous determination of their AGN types. In this paper, we treat them as radio galaxies. For many of them, we give their radio morphological FR type from the literature in Table \ref{sample1}.  Although typically FR-Is are associated with lower luminosities, and FR-IIs with higher luminosities, recent work has indicated that this is not the case, and that many low and high luminosity objects can be found with both morphological types \citep{Ming19}.  

\begin{figure}[t]
\begin{center}
\includegraphics[width=8cm]{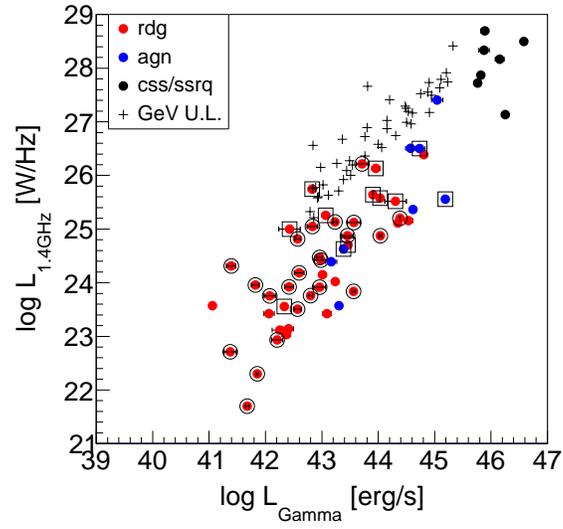}
\end{center}
\vspace*{0.3cm}
\caption{Gamma-ray luminosity (0.1--300 GeV)  vs radio luminosity (1.4 GHz). Red, blue, and black circles represent {\tt rdg}, {\tt agn}, and {\tt css/ssrq}, respectively. Data enclosed by circles or squares represent FR-I or FR-II, respectively. Crosses represent radio galaxies not detected by {\it Fermi}/LAT in \citet{Ming14}, where upper limits of GeV gamma-ray luminosity are plotted. }
\label{lglr}
\end{figure}

Figure \ref{lglr} shows a relationship between GeV gamma-ray and radio luminosity for our sample. We also plot radio galaxies not detected by {\it Fermi}/LAT from the radio flux-limited sample \citep{Ming14}, where we give upper limits on the GeV gamma-ray luminosity for an upper limit flux of $10^{-12}$ erg s$^{-1}$ cm$^{-2}$ (0.1--300 GeV). Interestingly, radio galaxies in the radio flux-limited sample are dominated by high luminosity objects (most are FR-II), while our GeV gamma-ray flux-limited sample contains many low luminosity ones. We will discuss this issue in \S\ref{sec:pop}.

Table~\ref{sample1} also summarizes the availability of the six spectral bands in 4FGL-DR2 and the flag for whether the source latitude $b$ is $|b|<20^\circ$ or not, since the Galactic diffuse emission can bring significant systematic uncertainties. We use these pieces of information in the LF modeling described later. Some objects have a detection flag at one or two energy bands since some energy bands (1--3, 3--10, and 10--30 GeV) have better sensitivity than others.

\subsection{X-ray data}
\label{sec:xray}

X-ray spectral information, combined with GeV gamma-rays, is also essential to investigating the high energy emission from MAGNs. We searched available X-ray observational data on the 4FGL-DR2 MAGNs using the {\it XMM-Newton}/EPIC \citep{Turn01}, {\it Chandra/ACIS} \citep{Weis00}, and {\it Swift}/XRT \citep{Gehr04,Burr05} archival lists. When available, we chose the data having the highest photon statistics for each MAGN. Since multiple X-ray satellite data sets are available for some of MAGNs, we selected the data in the order of priority: first {\it XMM-Newton}/EPIC, then {\it Chandra}/ACIS, and finally {\it Swift}/XRT. Then, we analyzed 20 data sets from {\it XMM-Newton}/EPIC, 15 data sets from {\it Chandra}/ACIS, and 9 data sets from {\it Swift}/XRT. During most of these observations the instruments were pointed directly at the 4FGL-DR2 MAGNs; however some were observed off-axis. For the 7 gamma-ray bright radio galaxies, we refer to the results of {\tt Suzaku} data analysis in \citet{Fuka15}, instead of analyzing other satellite data. As a result, we obtained X-ray data on 51 galaxies.
AGNs often show time variability in the X-ray band, but we confirmed that the variability amplitude is at most a factor of 2--3 by using {\it Swift}/XRT data for about 20 objects. Our results presented here are not affected significantly by such time variability.

{\it XMM-Newton}/EPIC data were analyzed with {\tt SAS} version 15.0.0 in the standard way; reprocessing, elimination of background flaring time region were done before making spectra. Photons within 60~arcsec of the object were extracted for spectral analysis, and background spectra were extracted in the annular region with radii from 500 to 550 arcsec. {\it Chandra}/ACIS data were analyzed with {\tt CIAO} version 4.11 in the standard way, using {\tt chandra\_repro} and {\tt specextract} for reprocessing and extraction of spectra, respectively. Photons within 2~arcsec of the object were extracted for spectral analysis. For some objects observed in the off-axis position, we set a larger extraction radius to collect enough photons. {\it Swift}/XRT data were retrieved from the UK Swift Science Data Center and  analyzed with {\tt HEASoft} version 6.19 in the standard way, using {\tt xselect}. Photons within 30~arcsec of the object were extracted for spectral analysis, and background spectra were extracted in the annular region with radii from 210 to 250 arcsec.

The X-ray spectra obtained above were fitted with {\tt XSPEC} version 12.9.0o. We adopt a spectral model of a single power-law with the Galactic interstellar medium absorption, with Hydrogen column density from \citet{Dick90}. Thermal emission components also appear in the soft X-ray band in some MAGNs (IC 1531, NGC 315, NGC 1316, NGC 2484, and NGC 4261), which originate in the hot interstellar or intra-group medium. In such cases, the thermal plasma model {\tt apec} \citep{Smit01} was added in the analysis to improve the goodness of fitting. Two objects (NGC 4261 and NGC 6251) have an excess absorption column density well above the Milky Way value, so we added an additional absorption component.  Table~\ref{sample2} summarizes the resulting X-ray photon indices and fluxes of our sample galaxies. Detailed results of X-ray spectral analysis will be presented in a separate paper.
 
Pile-up effects were non-negligible for five {\it Chandra}/ACIS selected MAGNs; NGC~2329, 3C~138, 3C~303, 3C~380, and NGC~3078. For these data, we fitted the spectra with {\tt Sherpa} by including the pile-up model {\tt jdpileup} to obtain the spectral parameters. The pile-up fraction is found to be 0.05--0.15 for these MAGNs, except for 3C~303. Since the {\it Chandra} data of 3C~303 were affected by significant pile-up, instead we analyzed the {\it NuStar} \citep{Harr13} data for this object.

For MAGNs, which are not significantly detected in the above X-ray data or have no X-ray pointing observations, we used the X-ray fluxes from the Rosat All-Sky Survey \citep[RASS;][]{Voge99} and XMM-Newton Slew Survey \citep[XSS;][]{Saxt08}. Their fluxes are also summarized in Table \ref{sample2}. The fluxes of five MAGNs are available in RASS, and one is available in XSS. As a result, X-ray information is available for 57 MAGNs. We set an upper limit on X-ray flux to $3\times10^{-13}$ erg s cm$^{-2}$ for the other 5 galaxies, based on the RASS survey limit.

\begin{figure}[t]
\begin{tabular}{cc}
\begin{minipage}{0.5\hsize}
\begin{center}
\includegraphics[width=8cm]{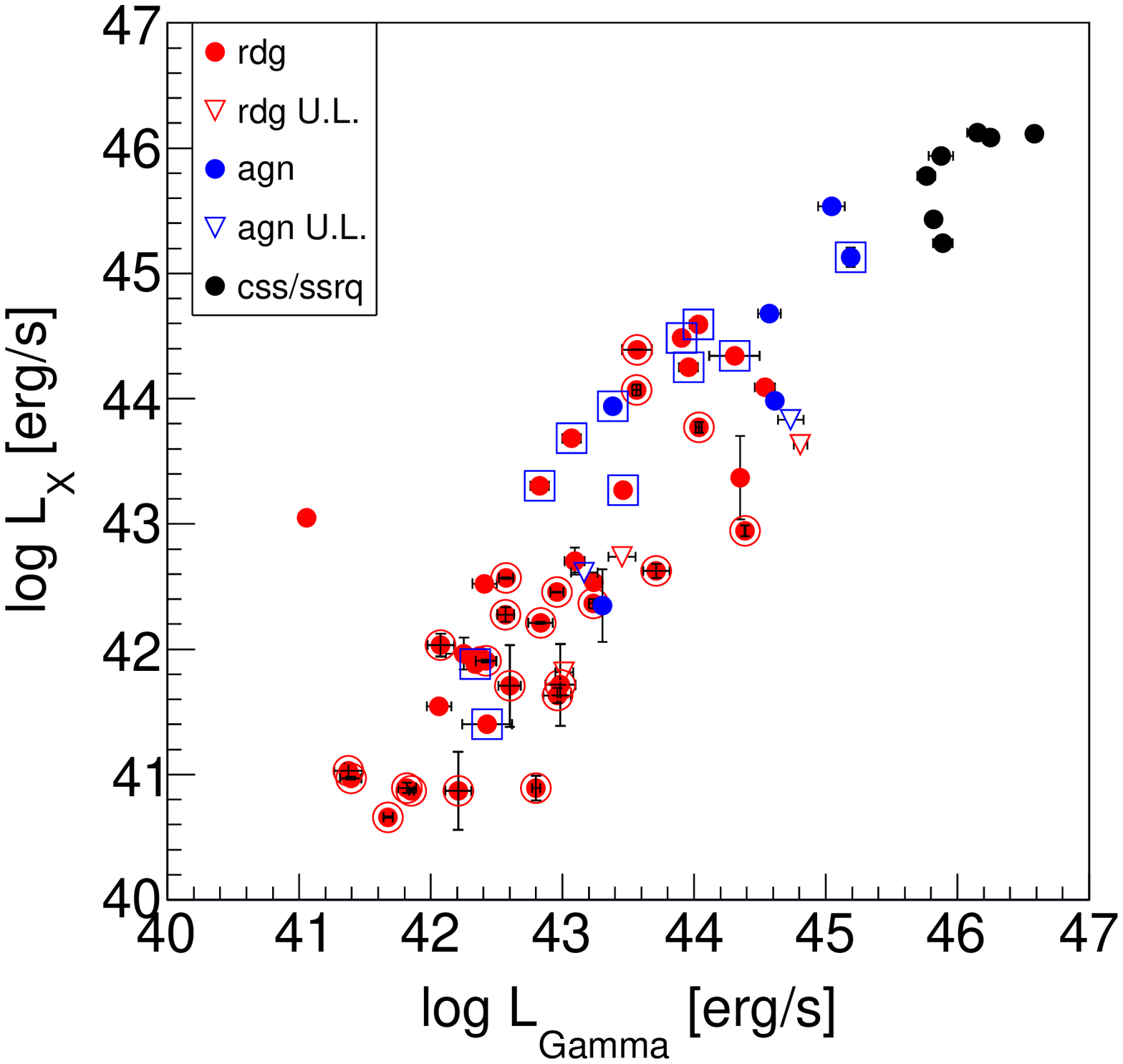}
\end{center}
\end{minipage}
\begin{minipage}{0.5\hsize}
\begin{center}
\includegraphics[width=8cm]{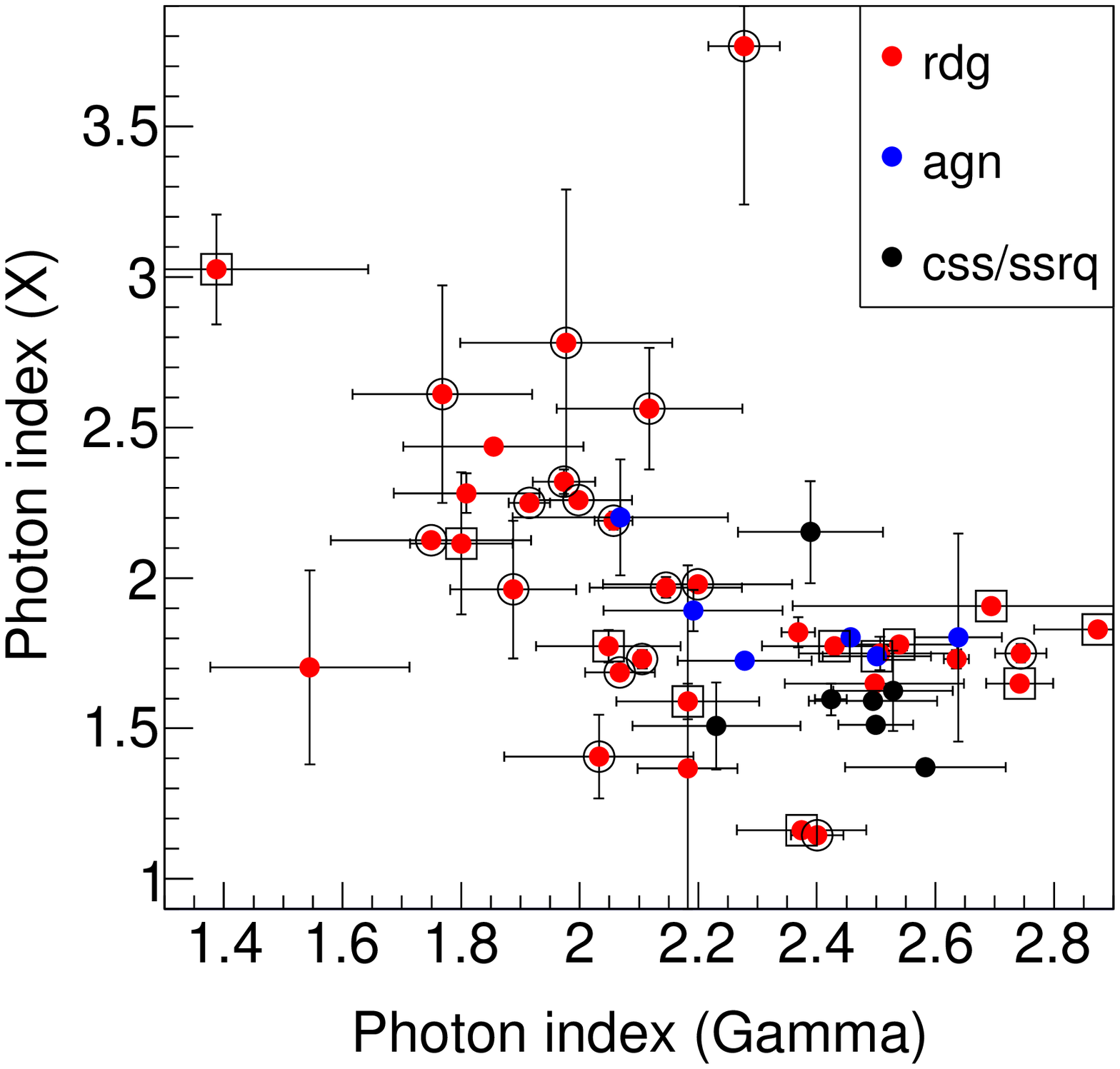}
\vspace*{0.5cm}
\end{center}
\end{minipage}
\end{tabular}
\vspace*{0.3cm}
\caption{{\it Left}: Gamma-ray luminosity (0.1--300 GeV)  vs X-ray luminosity (2--10 keV). {\it Right}: Gamma-ray photon index vs X-ray photon index. Red, blue, and black circles represent {\tt rdg}, {\tt agn}, and {\tt css/ssrq}, respectively. Data enclosed by circles or squares represent FR-I or FR-II, respectively. Triangles represent upper limits of X-ray luminosity. Objects with upper limints of X-ray luminosity are not shown in the right panel.}
\label{lglx}
\end{figure}

The left panel of Figure~\ref{lglx} shows the relation between the GeV gamma-ray luminosity and X-ray luminosity for the 4FGL-DR2 MAGNs. 
The X-ray luminosity $L_{\rm X}$ seems to correlate with the GeV gamma-ray luminosity $L_{\rm GeV}$. 
But, this correlation could be due to the mutual dependence on redshift and a selection bias that only gamma-ray emitting radio galaxies are plotted. We will discuss this issue in \S~\ref{sec:pop} and \S~\ref{sec:sed}.
FR-Is are located at the lower luminosity regime, while
CSSs and SSRQs are located at the highest luminosity end. 
FR-IIs have luminosities located in the middle, but no clear separation with FR-Is are seen in this plot, regardless of a clear separation in radio luminosity. Note that the one outlier with  low gamma-ray and high X-ray luminosity is Cen A.

The right panel of Figure \ref{lglx} shows the photon index relation between the GeV gamma-ray band and the X-ray band. MAGNs can be divided into two classes; one class has a soft GeV gamma-ray photon index of $\gtrsim2$ with a hard X-ray photon index of $\lesssim2$, and another has the opposite. The former group tends to contain higher-luminosity galaxies such as CSSs and SSRQs, while the latter tends to contain lower-luminosity galaxies.  This relation between X-ray and GeV gamma-ray photon indices is similar to that of blazars \citep{Samb10}.
There are two outliers. One is B2 1447+27 which has a hard X-ray photon index (1.70) and a hard gamma-ray photon index (1.54). The quality of the {\it Swift}/XRT X-ray spectrum is not good, and we cannot rule out that this spectrum is affected by excess absorption or other emission components.
The other one is NGC 2894, which has a very soft X-ray photon index (3.77) and a soft gamma-ray photon index (2.28).
This {\it Chandra}/ACIS X-ray spectrum is also not of good quality, and some absorption or additional components might affect the power-law model parameters.

\section{Spectral Energy Distribution from X-ray to GeV gamma-ray}
\label{ana:sed}

We can derive the SED from the X-ray to gamma-ray band by simultaneously fitting the X-ray spectral data and the six flux band measurements from {\it Fermi}-LAT. We adopt the following polynomial function so as to represent a flatter slope in the X-ray band and a steep drop in the highest energy regime: 
\begin{equation}
	\log{\nu F(\nu)}=A\left(\log{\nu}-\log{\nu_0}\right)^2\left(2\log{\nu}+\log{\nu_0}-3\log{\nu_1}\right)+B,
\end{equation}
where $F$ and $\nu$ are the energy flux and frequency, respectively, and $\nu_0$, $\nu_1$, $A$, and $B$ are fitting parameters. The parameter $A$ is negative, and $\nu_0$ and $\nu_1$ are limited to be in the range of $10^{16}$--$10^{17}$ Hz and $>10^{17}$ Hz, respectively. This function has a flux peak at $\nu_1$ and the differential coefficient becomes zero at $\nu_0$. Four free parameters were obtained by maximizing a log-likelihood $\ln{L}=-\sum_i \left(F(\nu_i)-F_i\right)^2/\left(2\delta F_i^2\right)$ with the Markov Chain Monte-Carlo (MCMC) technique, where $\nu_i$, $F_i$, and $\delta F_i$ are the measured frequency, flux, and flux error, respectively. 

When the X-ray photon index is larger than 2, the X-ray emission could be synchrotron radiation \citep{Fuka15}, and thus it is not appropriate to describe both X-ray and GeV gamma-ray with this function. In such a case, $-1$ or $10^9$ is added to the log-likelihood when the model is smaller or larger than the data flux, respectively, so that the model does not exceed the data; in other words, we use X-ray data as an upper limit for the high-energy emission component.
When there is no available X-ray data, we use an upper limit flux of $10^{-11}\ {\rm erg\ s^{-1}\ cm^{-2}}$ at 1~keV and fit in the same way as above so that the model does not exceed this flux. If only RASS or XSS data are available, we treat that flux as an upper flux at 1 keV, as above.

\begin{figure}[t]
\begin{tabular}{cc}
\begin{minipage}{0.5\hsize}
\begin{center}
\includegraphics[width=8cm]{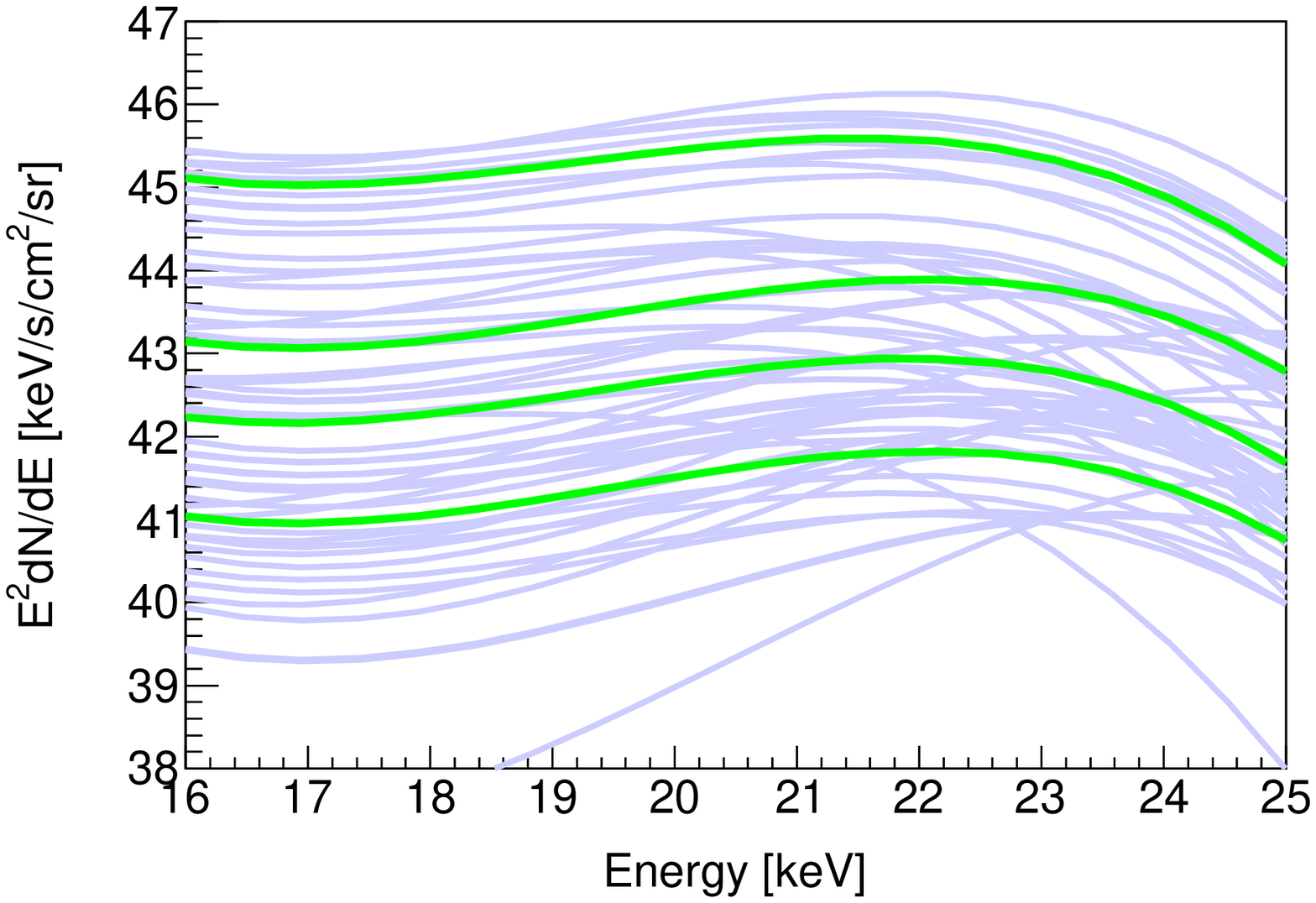}
\end{center}
\end{minipage}
\begin{minipage}{0.5\hsize}
\begin{center}
\includegraphics[width=8cm]{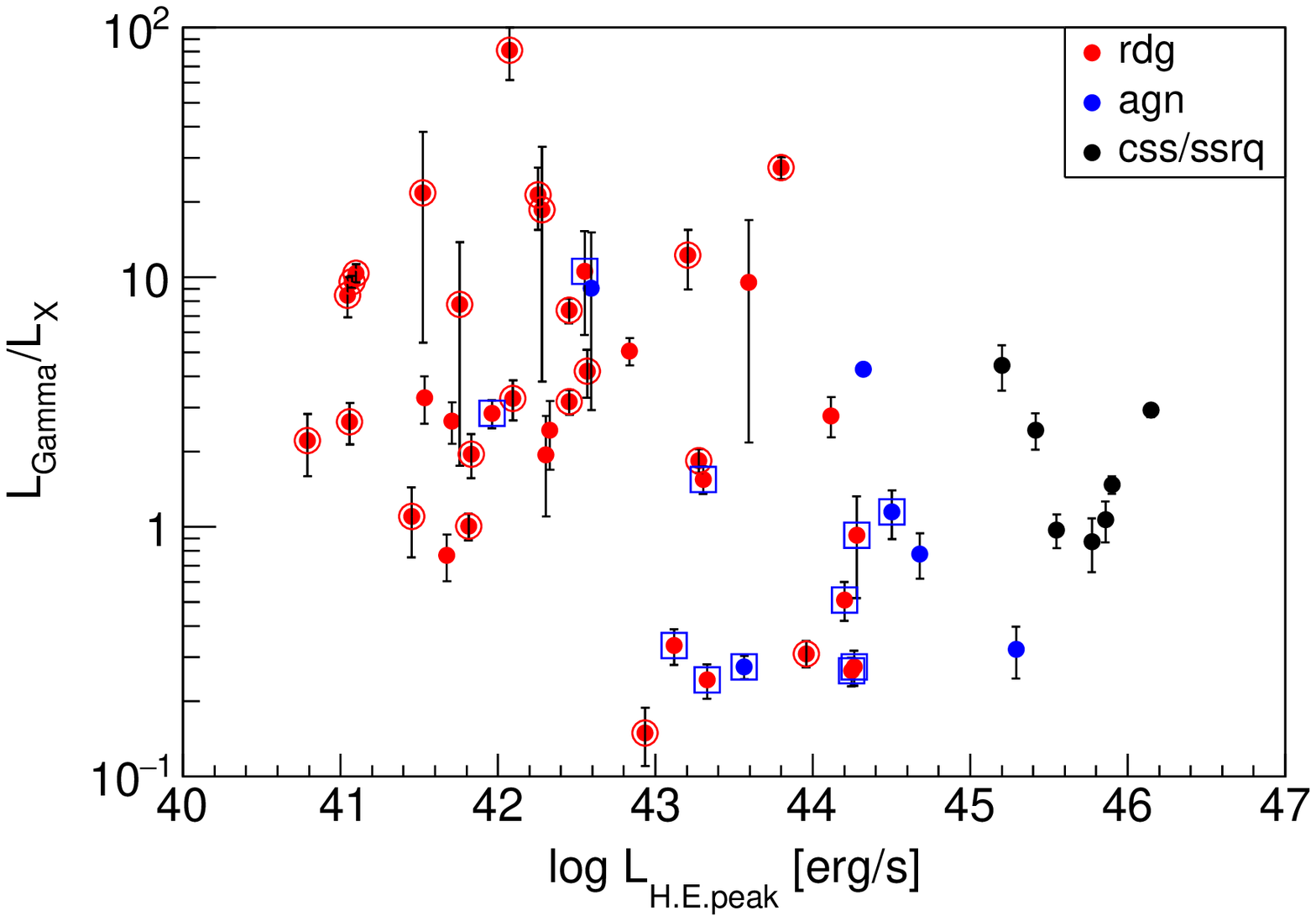}
\vspace*{0.5cm}
\end{center}
\end{minipage}
\end{tabular}
\vspace*{0.3cm}
\caption{Left: Dark blue line represents a model curve determined by fitting SED from X-ray to gamma-ray band for each galaxy. Thin green ;ome represents an average model curve in four luminosity ranges of $10^{40-42}$, $10^{42-43}$, $10^{43-44}$, $10^{44-46}$ erg s$^{-1}$ in the 1--3 GeV band. 
Right: SED Peak luminosity of high-energy component vs gamma-ray to X-ray luminosity ratio. Red, blue, and black circles represent {\tt rdg}, {\tt agn}, and {\tt css/ssrq}, respectively. Data enclosed by circles or squares represent FR-I or FR-II, respectively. 
 }
\label{modelcurve}
\end{figure}

Figure~\ref{modelcurve} left shows the SED curves obtained from X-rays to gamma-rays for our sample of MAGNs.
SED curves seem to depend on the luminosity, which is clearly seen as a dependence of a gamma-ray to X-ray luminosity ratio on the SED peak luminosity shown in Figure~\ref{modelcurve} right.
Table~\ref{sample2} summarizes the estimated peak frequencies and luminosities of the high energy components of MAGNs. Note that the high energy component in the X-ray band has contributions not only from jet emission but also from coronal emission.  Synchrotron peak frequencies and luminosities from jet emission are also interesting and have been studied in the past \citep{Fuka15,Keen20}, and thus we quote these values from 4LAC for our MAGNs.
For comparison, we also quote the values of blazars. Synchrotron peak values are taken from 4LAC, while IC peak values are from \citet{1SED}, where the peak was estimated by fitting the available SED data with polynomial functions.
Note that these peak values are obtained by just fitting SEDs by polynomial functions without modeling non-jet components in our study. For the 4LAC paper \citep{4LAC}, if non-jet components are clearly seen in the SED, the data affected by such components are ignored in the fitting.
However, for MAGNs, non-jet components are not clearly seen in the SED.
Therefore, these peak frequencies and luminosities could be affected by non-jet emission components.

Figure~\ref{sedpeak} shows the relation between peak frequency and peak luminosity for the synchrotron and high energy components.
It can be seen from the top-left figure that the
synchrotron peak frequency of radio galaxies has a wide distribution, as for blazars, from infrared to X-rays bands, but tends to concentrate towards a lower frequency. However, we note that the infrared, optical, and X-ray bands could be significantly contaminated by the emission from host galaxies and AGN disks/coronae and thus the SED peak position could be affected.
There is a weak correlation between the peak frequencies of the synchrotron and high energy components, but a scatter could again be attributed to the contamination from other emission components.
The highest synchrotron peak frequency around $10^{17.5}$ Hz is TXS 1516+064, which has a flat GeV gamma-ray photon index of 1.75.

The top-right panel shows that 
the peak luminosity of the synchrotron and high energy components have a tight correlation
in radio galaxies as well as blazars.
Note that the peak luminosity of the high energy component of radio galaxies
is somewhat lower than that of the synchrotron component; this is contrast to the Compton dominance of high-luminosity blazars, especially FSRQs.
Contamination of other emission components to the synchrotron peak luminosity makes this trend less certain.  These properties will be discussed again in \S~\ref{sec:pop}.

The bottom-left panel shows that radio galaxies do not seem to follow the anticorrelation between synchrotron peak frequency and luminosity sometimes called the blazar sequence \citep{Fossati1998, Kubo1998}. 
This could also be caused by contamination of non-jet emission components.

The bottom-right panel shows that the high energy component has an anti-correlation between peak frequency and luminosity for radio galaxies that is similar to blazars.
The wide range of peak frequencies is also similar to that of blazars.
Note that the outlier object with a low peak frequency of $10^{19.7}$ Hz and a low luminosity of $10^{41.5}$ erg s$^{-1}$  is Cen A.
This trend is consistent with the right panel of Fig.~\ref{lglx}. This feature can be understood as follows. When the GeV gamma-ray spectrum is soft, implying the GeV band is from the high-energy tail of the inverse Compton scattering (IC) component, the X-ray emission corresponds to the low-energy tail of the IC component, which will be hard; or the X-rays could be from coronal emission, with typical X-ray photon index is around 1.7--1.9 in Seyfert galaxies \citep{Nand94}. When the GeV gamma-ray spectrum is hard, implying that the GeV emission comes from the low-energy tail of the inverse Compton (IC) component, the X-ray emission corresponds to the high-energy tail of the synchrotron radiation, producing a soft spectrum.

\section{GeV gamma-ray luminosity function}
\label{sec:glf}

\subsection{Trial luminosity function}

A luminosity function represents a source number density as a function of luminosity and redshift. The luminosity-Dependent Density Evolution (LDDE) model is often used for radio-quiet AGNs and blazars \citep{Ueda2003, Naru06, Ueda14, Ajel15}. It is given by

\begin{eqnarray}
  \Phi(L_\gamma,z)&=&\frac{d^2N}{dzdL_X}\\ \nonumber
  &=&\frac{A}{\ln(10)L_\gamma}
  \left[\left(\frac{L_\gamma}{L_*}\right)^{\gamma_1} +
   \left(\frac{L_\gamma}{L_*}\right)^{\gamma_2}  \right]^{-1}\\ 
   &\times&
  \left[\left(\frac{1+z}{1+z_c(L_\gamma)}\right)^{p_1^*} +
   \left(\frac{1+z}{1+z_c(L_\gamma)}\right)^{p_2^*}    \right]^{-1} ,
\end{eqnarray}

where $z_c(L_\gamma)=z_c({L_\gamma}/{L_z})^{\alpha}$. Here $z$ and $L_\gamma$ are the redshift and gamma-ray luminosity in a certain energy band, respectively. We assume $L_z=10^{42.5}~ {\rm erg\ s^{-1}}$. The model parameters are a normalization constant $A$, a characteristic gamma-ray luminosity $L_*$, a characteristic redshift $z_c$, two luminosity indices $\gamma_1$ and $\gamma_2$, two redshift indices $p_1^*$ and $p_2^*$, and $\alpha$. We adopt
\begin{eqnarray}
	p_1^*&=&p_1+\tau\left(\log_{10} L_\gamma-\log_{10}L_p\right)\ 
\end{eqnarray}
and
\begin{eqnarray}
	p_2^*&=&p_2+\delta\left(\log_{10}L_\gamma-\log_{10}L_p\right),
\end{eqnarray}
which were used for the GLF of blazars \citep{Ajel15}. We set $L_p =10^{42.5}~ {\rm erg\ s^{-1}}$.

Since our MAGN sample lacks high redshift objects, we cannot determine the redshift dependence precisely. Therefore, we apply eight previously established redshift evolution models of AGNs from the literature. Models {\tt BLLz} and {\tt FSRQz} assume the indices $p_1$ and $p_2$ to be the same as those of GLFs of BL Lacs (LDDE$_1$ in \citet{Ajel14} and FSRQs \citep{Ajel12}, respectively, with a free $z_c$ and a single power-law with a cut off at higher luminosity for luminosity dependence. We fix $L_*=10^{46}~ {\rm erg\ s^{-1}}$, higher than the luminosity in each band of our sample galaxies, and $\gamma_2$ is fixed to 5.0. These two LFs assume $p_1^*$ and $p_2^*$ do not depend on the luminosity ($\tau=\delta=0$). Models {\tt BLLp2} and {\tt FSRQp2} assume the redshift dependence of $p_1$ and $z_c$  to be the same as those of GLFs of BL Lacs \citep[LDDE$_1$,][]{Ajel14} and FSRQs \citep{Ajel12}, respectively, with a low-redshift index $p_2$ to be free and the same luminosity dependence as {\tt BLLz} and {\tt FSRQz}. Model {\tt BLL0}, {\tt FSRQ0}, and {\tt BLAZAR0} assume the same LF shape as that of GLFs of BL Lacs \citep[LDDE$_2$,][]{Ajel14}, FSRQs \citep{Ajel12}, and blazars \citep[FSRQ + BL Lac,][]{Ajel15}, respectively; parameters other than $A$ and $L_*$ are fixed to those from these papers. Thus, the luminosity dependence of the LF is also the same as that of blazars by assumption.
The last model {\tt RLF0} assumes a similar redshift dependence to that of the radio galaxy LF obtained in the radio band \citep{Will01}, which was used in the previous study of GLF of radio galaxies \citep{Inou11,DiMa14}. It assumes a single power-law with an exponential cut off for luminosity dependence with the index as a free parameter. See Table~\ref{lf1} for details on fixed and free parameters.

\subsection{Parameter Constraint}

Parameters of the trial luminosity function $\Phi(L_\gamma,z)$ are determined by the likelihood analysis.
In order to maximize the log-likelihood $\ln{\mathcal{L}}$, a
MCMC method using the adaptive Metropolis algorithm \citep{Haar01} is applied, following \citet{Yama20}.
The log-likelihood function is defined as

\begin{eqnarray} \nonumber
 \ln \mathcal{L} &=& \sum^{N_{\rm obs}}_i \ln \left( \Phi (L_{\gamma,i},z_i)  \right)\\
   &-&  \int_{z_{\rm min}}^{z_{\rm max}} dz
  \int_{L_{\gamma, \rm min}}^{L_{\gamma, \rm max}} dL_\gamma \Phi(L_\gamma,z)S(L_\gamma,z),
\label{Lfanction}
\end{eqnarray}
where $S(L_\gamma,z)$ is the sky coverage function described below. We set $z_{\rm min}=0$, $z_{\rm max}=6$, $L_{\gamma, \rm min}=2\times10^{40}~ {\rm erg\ s^{-1}}$, and $L_{\gamma, \rm max}=10^{46}~ {\rm erg\ s^{-1}}$. The redshift and luminosity ranges are based on the observed ranges. 

\begin{figure}[t]
\includegraphics[width=8cm]{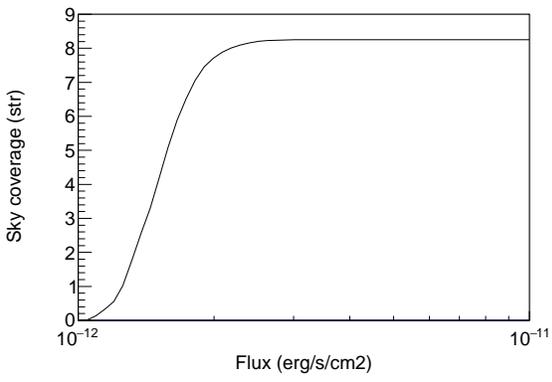}
\vspace*{0.3cm}
\caption{Sky coverage function at 0.1--100 GeV of {\it Fermi}/LAT 4FGL-DR2 catalog, created from the sensitivity map. }
\label{skycov}
\end{figure}

Sky coverage $S(L_\gamma, z)$ as a function of flux is calculated from the public 4FGL-DR2 sensitivity map\footnote{\url{https://fermi.gsfc.nasa.gov/ssc/data/access/lat/10yr_catalog/detthresh_P8R3_source_10years_PL22.fits}} at $|b|>20^{\circ}$.
This map is for the 0.1--100 GeV  band and was created assuming a single-power-law source energy spectrum with a photon index of 2.2 (Figure \ref{skycov}). 
At first we scaled it to the function for the 1--3 GeV energy band by using a power-law spectrum with a photon index of 2.2.
Then we scaled the function for the 1--3 GeV band to that for the energy bands of 0.1--0.3, 0.3--1.0, 3.0--10.0, 10.00-30.0, 30.00-300.0 GeV by the sensitivity ratio between 1--3 GeV and each band, where the {\it Fermi}/LAT sensitivity curve as a function of energy\footnote{\url{https://www.slac.stanford.edu/exp/glast/groups/canda/lat_Performance_files/differential_flux_sensitivity_p8r3_source_v2_all_10yr_zmax100_n10.0_e1.50_ts25.png};
a red curve of (l,b)=(0,30)} is referred to.

\subsection{Gamma-ray Luminosity Function based on the 1--3 GeV flux}

We constrain the LF using the  4FGL-DR2 1--3 GeV band flux. The {\it Fermi}/LAT has the best sensitivity in this band and we can use the largest number of detected MAGNs. We use MAGNs, which are detected in 1--3 GeV. We exclude MAGNs whose flux is lower than the 1--3 GeV detection limit where the sky coverage introduced in the previous subsection becomes 0.
We do not include CSS and SSRQ galaxies. Since they have a higher redshift and higher luminosity than others and their gamma-ray emission mechanisms are still veiled in mystery, they are a different population from other MAGNs. We also exclude MAGNs located near the Galactic plane at $|b|\leq20^{\circ}$. As a result, 34 radio galaxies are used for the GLF, where 7 {\tt AGN}s are included. Table~\ref{lf1} summarizes the parameters obtained for LFs for 8 trial models in this energy band.

Overall, the eight models reproduce the data with similar likelihood values, but
{\tt BLL0} and {\tt RLF0} give a significantly lower likelihood than others. 
{\tt BLLz} and {\tt FSRQz} give small $z_c = 0.001$, and {\tt BLLsp2} and {\tt FSRQsp2} give a positive $p_2$, meaning that these four models give a negative evolution. A luminosity-dependent slope is obtained to be $\gamma_2=1.3-1.5$. {\tt BLL0}, {\tt FSRQ0}, and {\tt BLAZAR0} give a characteristic gamma-ray luminosity $L_*$ of $\sim10^{44}~ {\rm erg\ s^{-1}}$, $\sim10^{42}~ {\rm erg\ s^{-1}}$, and $\sim10^{43.4}~ {\rm erg\ s^{-1}}$, respectively, which are several orders of magnitude smaller than those of blazars ($10^{47-48}~{\rm erg\ s^{-1}}$). A smaller $L_*$ for {\tt FSRQ0} than the other two could be due to a flatter slope of the luminosity dependence, $\gamma_2=0.21$, of LF at higher luminosity.

In order to quantitatively look at how well the obtained LF matches the data, we calculate the observed redshift, luminosity, and cumulated source count (logN-logS) distributions, respectively, as follows:

\begin{eqnarray}
 \frac{dN}{dz} &=& 4\pi \int_{L_{\gamma, {\rm min}}}^{L_{\gamma, {\rm max}}} \Phi(L_\gamma,z)S(L_\gamma,z) dL_\gamma \frac{d^2V}{dzd\Omega}\\
\frac{dN}{dL_\gamma} &=& 4\pi
 \int_{z_{\rm min}}^{z_{\rm max}} \Phi(L_\gamma,z)S(L_\gamma,z)dz \frac{d^2V}{dzd\Omega} \\
 N(>S_0)&=&\int^{z_{\rm max}}_{z_{\rm min}} dz \frac{d^2V}{dzd\Omega}
  \int^{L_{\gamma, {\rm max}}}_{L_0}  \Phi(L_\gamma,z)dL_\gamma
\label{dNdzdNdL}
\end{eqnarray}

where ${d^2V}/{dzd\Omega}$ is the comoving volume per redshift per solid angle, $L_0 = 4\pi D_L(z)^2 S_0 (1+z)^{\Gamma -2}$, $D_L(z)$ is the luminosity distance, $S_0$ is the observed gamma-ray flux, and $\Gamma$ is the power-law photon index of the gamma-ray spectra. Here, we fix $\Gamma=2.25$, the mean value for our sample (see the next subsection).

\begin{figure}[t]
\begin{tabular}{cc}
\begin{minipage}{0.5\hsize}
\begin{center}
\includegraphics[width=8cm]{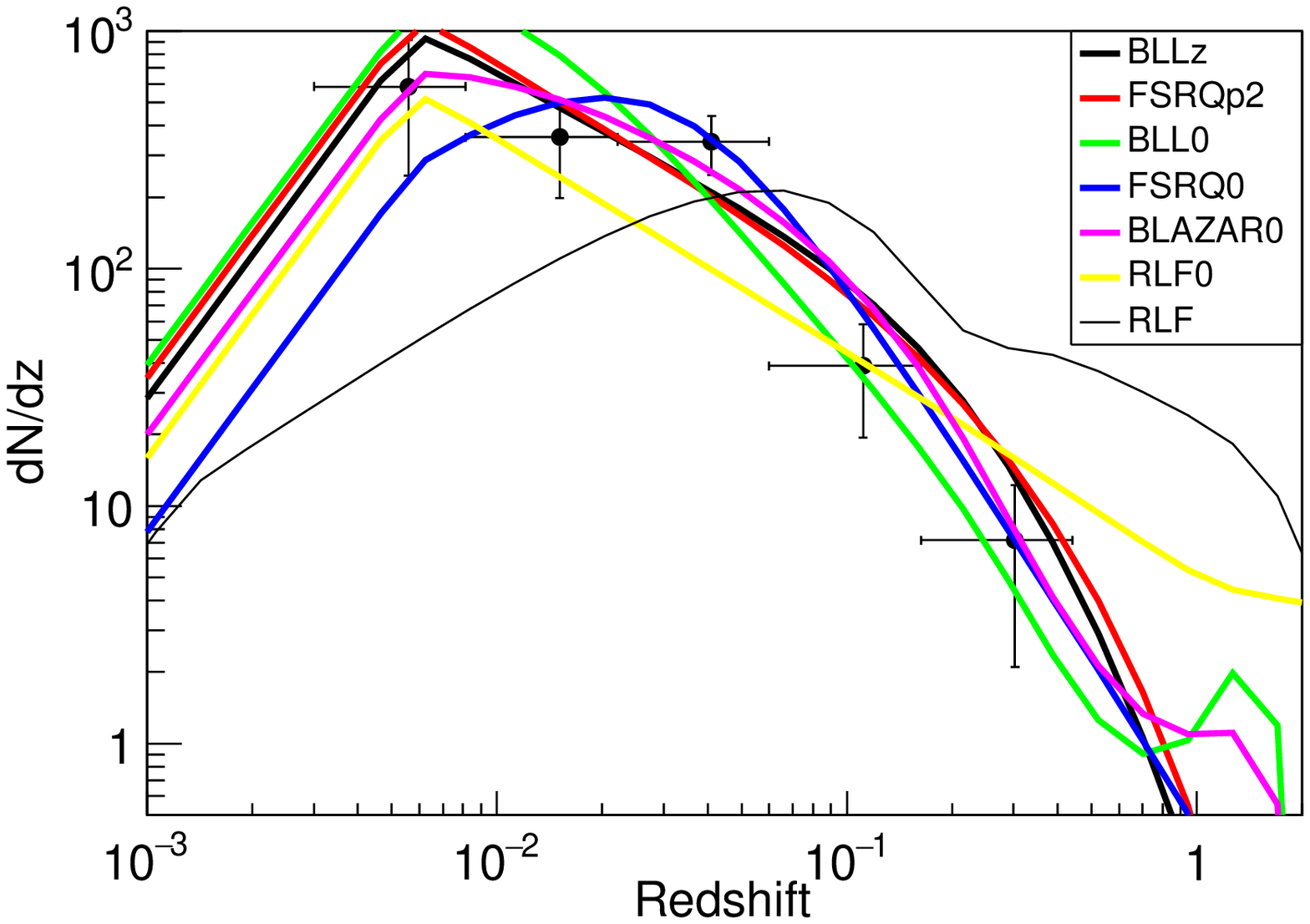}
\end{center}
\end{minipage}
\begin{minipage}{0.5\hsize}
\begin{center}
\includegraphics[width=8cm]{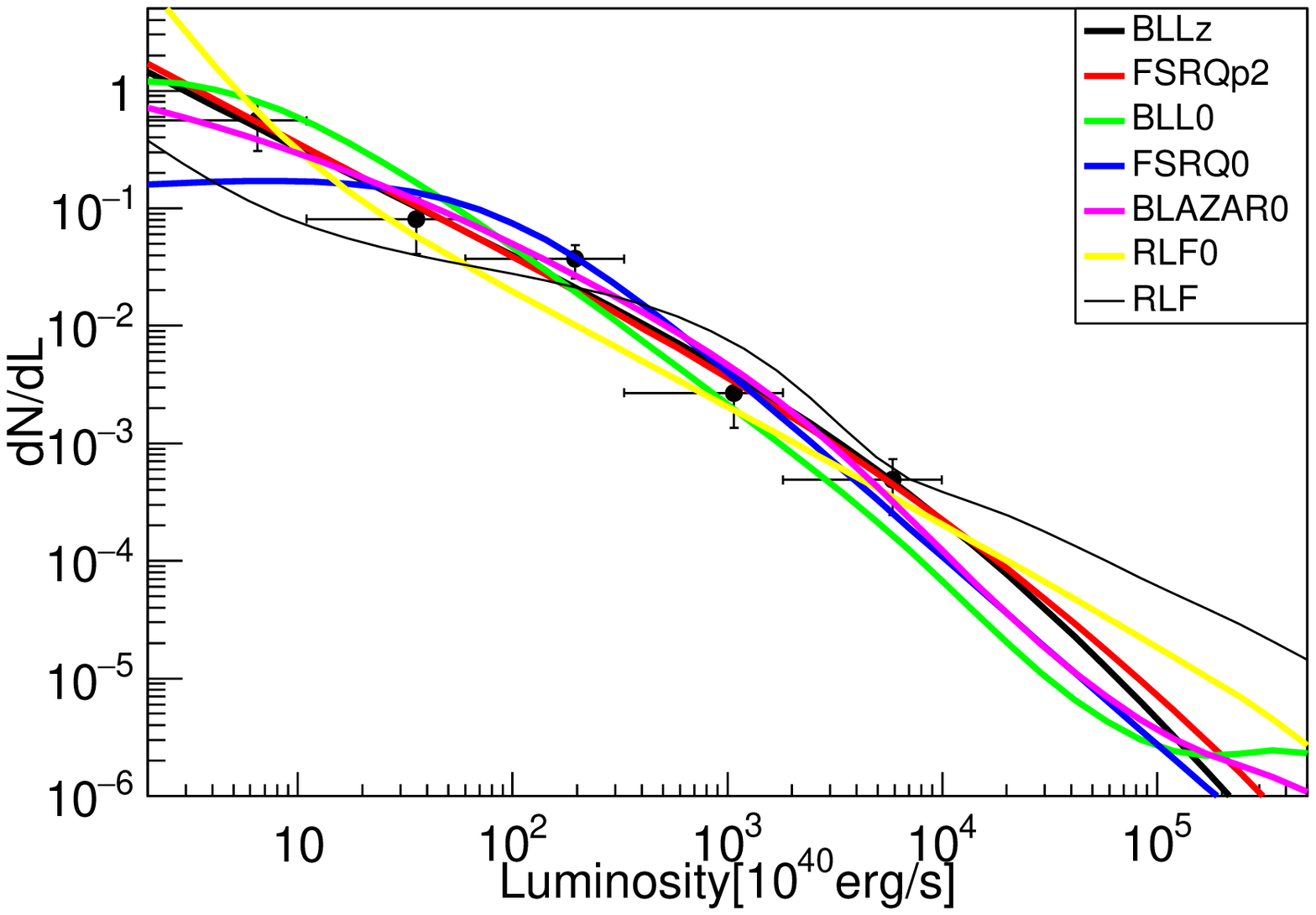}
\vspace*{0.5cm}
\end{center}
\end{minipage}\\
\begin{minipage}{0.5\hsize}
\begin{center}
\includegraphics[width=8cm]{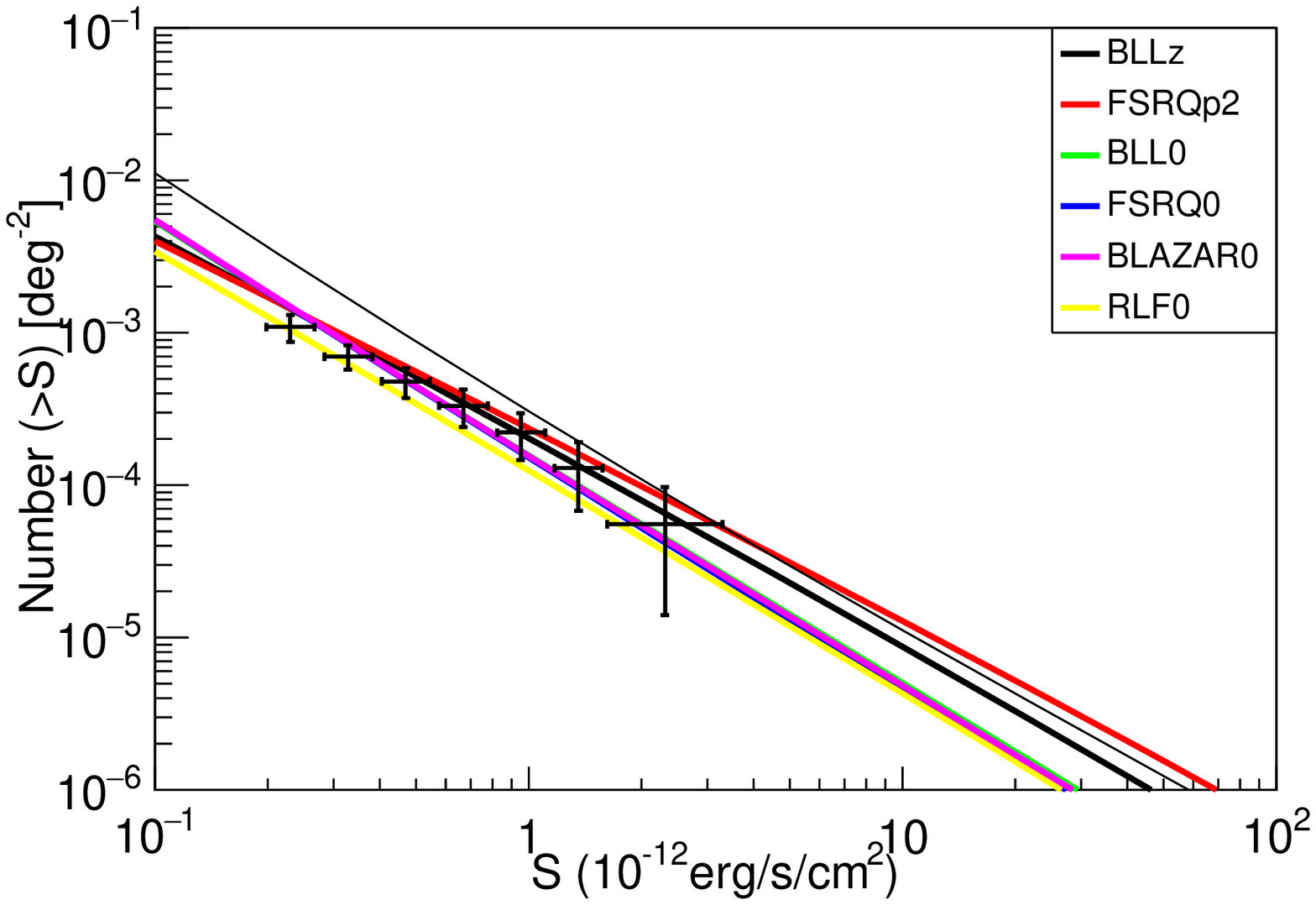}
\vspace*{0.5cm}
\end{center}
\end{minipage}
\end{tabular}
\vspace*{0.3cm}
\caption{{\it Left}: Redshift distribution of our sample, prediction from the best-fit LDDE models. {\it Middle} and {\it Right} are that for luminosity and cumulative source number count, respectively. Thick black, red, light green, blue, purple, and yellow represent a model curve of {\tt BLLz}, {\tt FSRQp2}, {\tt BLL0}, {\tt FSRQ0}, {\tt BLAZAR0}, and {\tt RLF0}, respectively. Thin black represents {\tt RLF}. The horizontal axis of the cumulative source number count is a gamma-ray flux in 1--3 GeV.}
\label{zldist}
\end{figure}

The three panels of Figure~\ref{zldist} show model curves together with the observed data points. We correct data points by the sky coverage factor for the logN-logS plot.  Hereafter, we do not show the model curves of {\tt FSRQz} and {\tt BLLp2}, since they give almost identical curves to that of {\tt BLLz}.

All the models reproduce the data distribution of redshift and luminosity well, but {\tt BLL0} and {\tt RLF0} models give a somewhat larger deviation from the data for the redshift and luminosity distribution.
We calculate the Kolmogorov-Smirnov (KS) probability \citep{Eadi06} for luminosity and redshift distributions for each model, shown in Table \ref{lf1}.
The {\tt BLL0} and {\tt RLF0} models show statistically different behavior comparing with the data based on the KS-test values obtained. Therefore, we reject these models.

Figure~\ref{zldist} also shows the model curve based on the LF used in \citet{Inou11} for comparison. This LF ({\tt RLF}) was obtained in the radio band for radio galaxies \citep{Will01}. For this RLF, we set $L_{\gamma, {\rm min}}=10^{39}$ erg s$^{-1}$ and $L_{\gamma, {\rm max}}=10^{48}$ erg s$^{-1}$, following \citet{Inou11}. As the KS-test values are small, $\sim0.01$, this LF does not adequately represent the gamma-ray luminosity function of radio galaxies. This indicates that GeV gamma-ray emitting radio galaxies are not the same population as radio emitting radio galaxies, as discussed in \S\ \ref{sec:pop}.

\begin{figure}[t]
\begin{tabular}{cc}
\begin{minipage}{0.3333\hsize}
\begin{center}
\includegraphics[width=5cm]{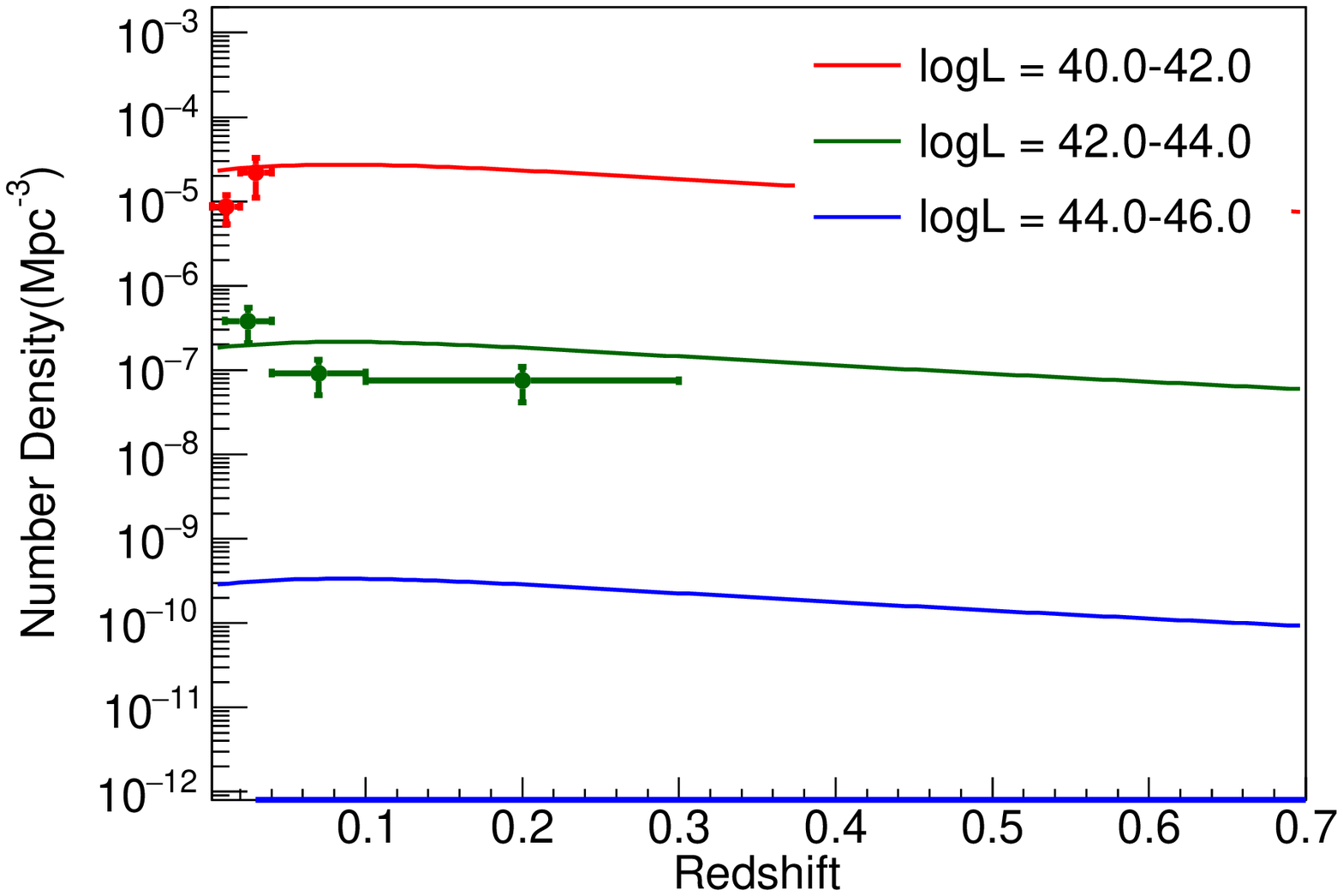}
\end{center}
\end{minipage}
\begin{minipage}{0.3333\hsize}
\begin{center}
\includegraphics[width=5cm]{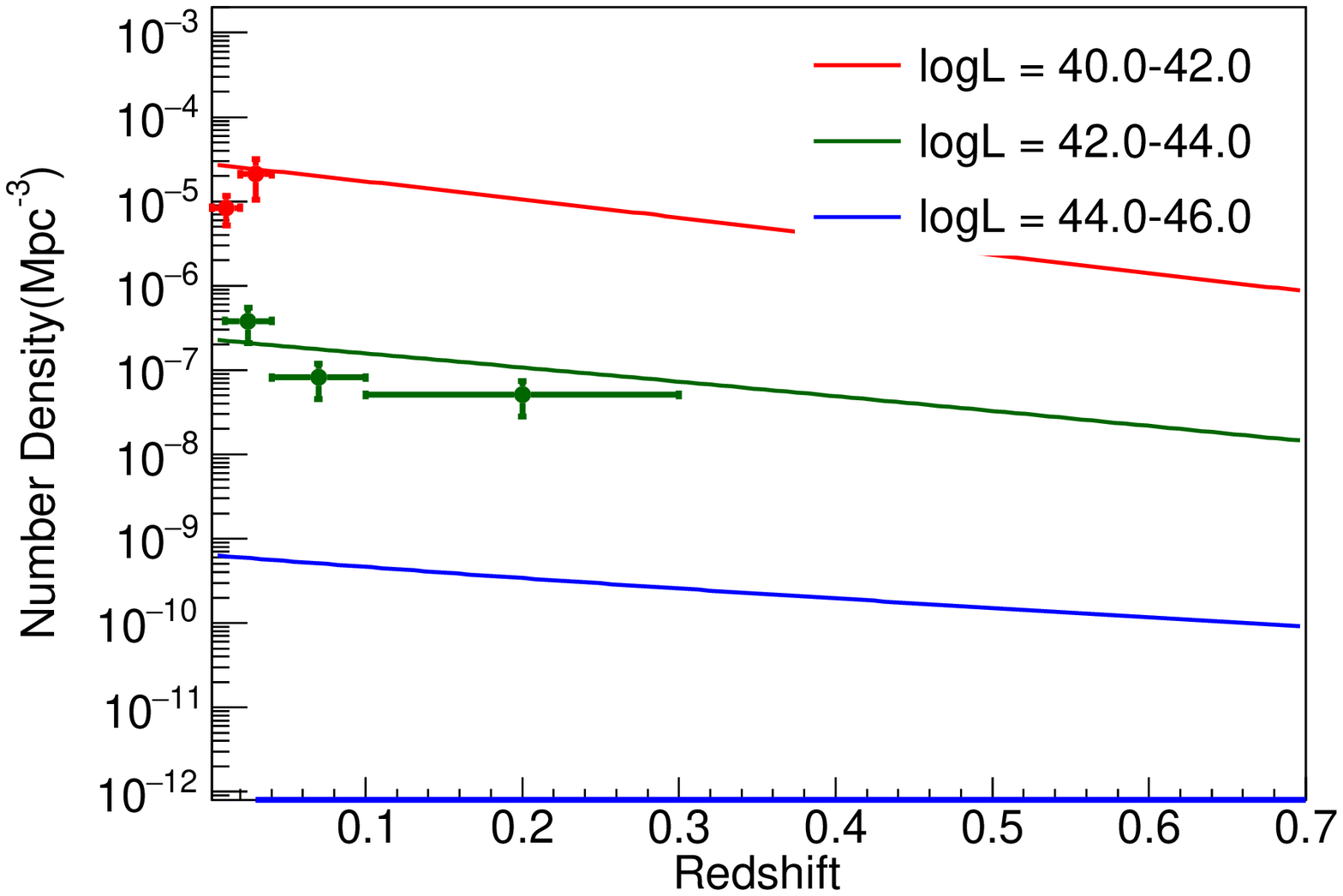}
\vspace*{0.5cm}
\end{center}
\end{minipage}
\begin{minipage}{0.3333\hsize}
\begin{center}
\includegraphics[width=5cm]{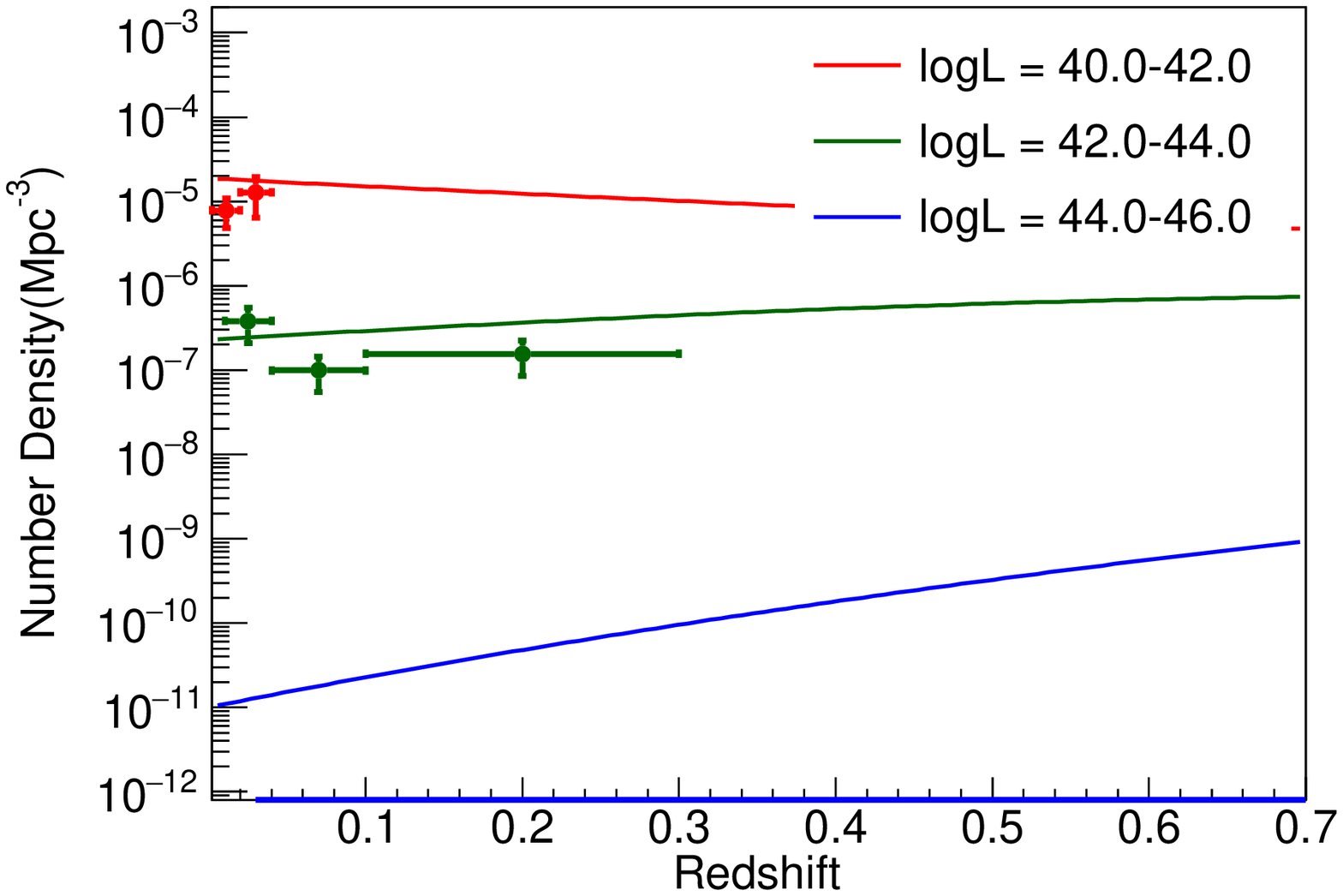}
\vspace*{0.5cm}
\end{center}
\end{minipage}
\end{tabular}
\vspace*{0.3cm}
\caption{Gamma-ray Luminosity function for the comoving number density of our sample in various luminosity bins. Data points are deconvolved by dividing them by $N^{\rm obs}/N^{\rm mdl}$. Model curves correspond to the best-fit LDDE models at different redshift bins. (\it Left): {\tt BLLz}. (\it Middle): {\tt FSRQp2}. (\it Right): {\tt BLAZAR0}. }
\label{lfzdist}
\end{figure}

For the logN-logS plot, {\tt BLLz} and {\tt FSRQp2} reproduce the data well while the other models tend to underestimate the number count at higher flux. Since these LFs have a strong positive evolution (as shown later in Figure \ref{lfzdist}), low-z radio galaxies are predicted not to be dominant in the number count, even at a higher flux. Therefore, we only consider the {\tt BLLz} and {\tt FSRQp2} models hereafter and show the model curves of only these models, together with the model curve of {\tt BLAZAR0} as a reference.

\begin{figure}[t]
\begin{tabular}{cc}
\begin{minipage}{0.33\hsize}
\begin{center}
\includegraphics[width=5cm]{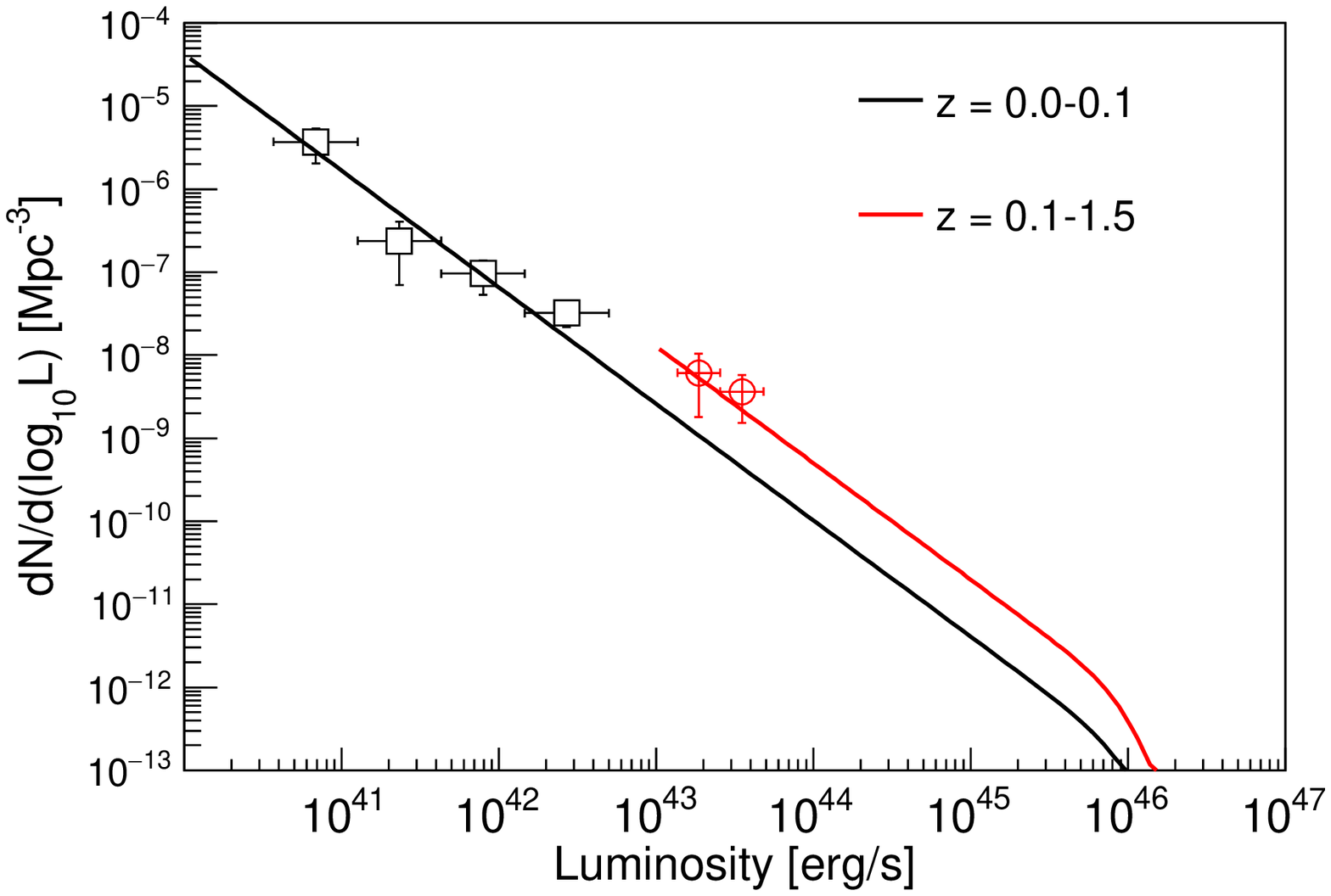}
\end{center}
\end{minipage}
\begin{minipage}{0.33\hsize}
\begin{center}
\includegraphics[width=5cm]{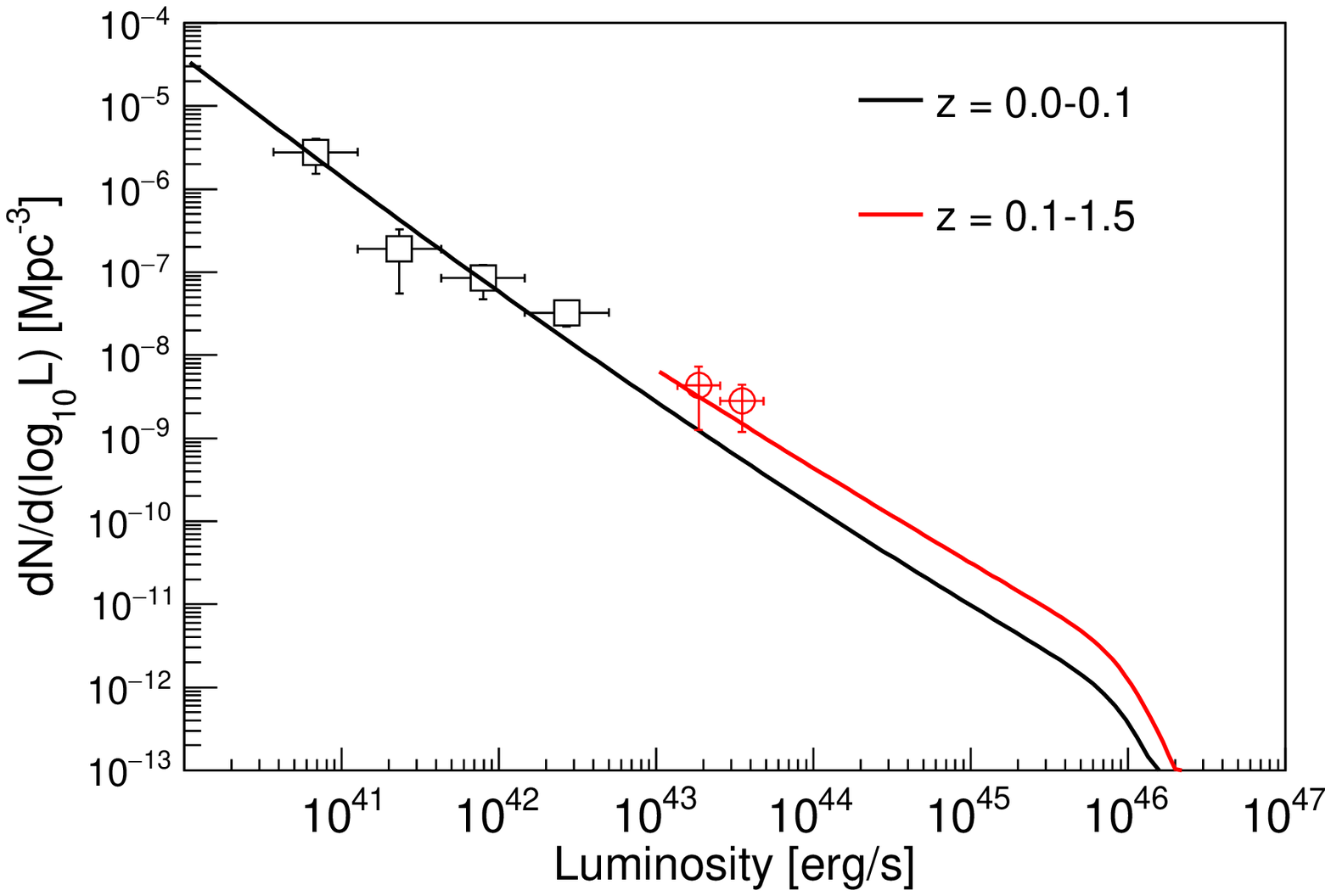}
\vspace*{0.5cm}
\end{center}
\end{minipage}
\begin{minipage}{0.33\hsize}
\begin{center}
\includegraphics[width=5cm]{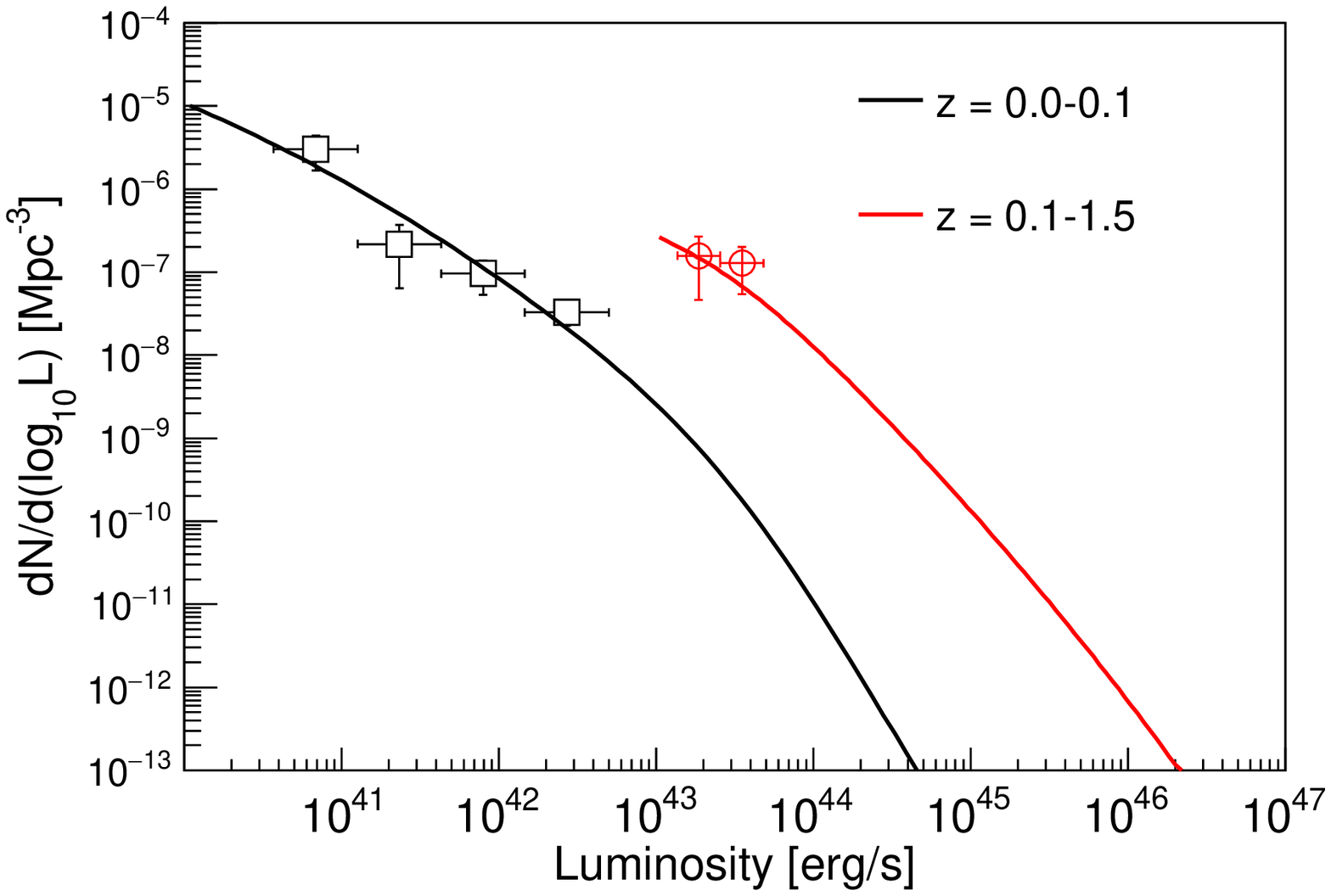}
\end{center}
\end{minipage}
\end{tabular}
\vspace*{0.3cm}
\caption{Gamma-ray luminosity function of our sample in various redshift bins. Model curves correspond to the best-fit LDDE models at different redshift bins. Data points are deconvolved by dividing them by $N^{\rm obs}/N^{\rm mdl}$. (\it Left): {\tt BLLz}. (\it Middle): {\tt FSRQp2}. (\it Right): {\tt BLAZAR0}. }
\label{lfldist}
\end{figure}

Figure \ref{lfldist} shows a visualization of LFs for two redshift regimes: $0.0\le z< 0.1$ and $0.1\le z<1.5$. We use the ``$N^{\rm obs}/N^{\rm mdl}$'' method  \citep{LaFr97,Miya01}. We deconvolve the observed data points by dividing them by $N_i^{\rm obs}/N_i^{\rm mdl}$, where $N_i^{\rm obs}$ and $N_i^{\rm mdl}$ are the observed and the predicted number of radio galaxies in that bin, respectively, and are calculated by integrating $\Phi(L_\gamma,z)S(L_\gamma,z)$ and $\Phi(L_\gamma,z)$, respectively, in that bin.
Note that {\tt BLLz} and {\tt FSRQp2} assume a single power-law luminosity dependence, while {\tt BLAZAR0} assumes a broken power-law luminosity dependence.
All models match the data well, and they would be distinguished if radio galaxies at various redshifts were sampled.
Figure \ref{lfzdist} visualizes the obtained LFs as a function of redshift for various luminosity ranges with the ``$N^{\rm obs}/N^{\rm mdl}$'' method.
Here we plot the LF in a redshift range where the observed radio galaxies exist.
Different redshift dependence among LFs is clearly seen.
In these figures, we cannot say which model matchs the data the best.
On the other hand, {\tt BLLz} and {\tt FSRQp2} seem to reproduce the logN-logS relation the best as described above (Figure \ref{zldist}).
This indicates that the gamma-ray emitting radio galaxies have a lower redshift peak in the LF than blazars and other AGNs.

\subsection{Gamma-ray Luminosity Function Based on the Six Energy Band Flux Measurements}
\label{EneLF}

We also constrain the GLF in each of six energy bands where the flux measurements are listed in 4FGL-DR2. In this case, we use the {\tt BLLz} model where free parameters are the same as in 1.0--3.0 GeV band; $A$, $\gamma_2$, and $z_c$. The sky coverage function for each of energy band is used in this fitting. Table~\ref{lf2} summarizes the fitting results.
The selection condition for MAGNs used in this analysis is the same as that for 1--3 GeV band.
The parameters $\gamma_2$ and $z_c$ are similar among all energy bands, but $A$ has a dependence on the energy band.

Next we try a common GLF among the 6 energy bands, by taking into account a distribution of gamma-ray photon indices $\Gamma$, $G(\Gamma)$.
The observed photon index has a wide range of values, 1.4--2.5, and could cause a difference in the GLF among the 6 energy bands as shown above.
In this case, the log-likelihood in the {\it j}-th energy band is defined as 

\begin{eqnarray} \nonumber
 \ln \mathcal{L}_j &=& \sum^{N_{\rm obs}}_i \ln \left( \Phi (L_{\gamma,i},z_i)G(\Gamma)  \right)\\
   &-&  \int_{z_{\rm min}}^{z_{\rm max}} dz
  \int_{L_{\gamma, \rm min}}^{L_{\gamma, \rm max}} dL_\gamma\int_{\Gamma_{\rm min}}^{\Gamma_{\rm max}}d\Gamma \Phi(L_\gamma,z)G(\Gamma)S_j(L_\gamma,z)
\label{Lfanction}
\end{eqnarray}

where $G(\Gamma)=G_0\exp[{-(\Gamma-\mu)^2}/{2\sigma^2}]$ and
$S_j(L_\gamma,z)$ is the sky coverage function in the {\it j}-th energy band.
$\mu$ and $\sigma$ are the mean and variance of the photon index distribution, respectively, and we set a lower and upper bounds to be $\Gamma_{\rm min}=1.3$ and $\Gamma_{\rm max}=3.0$, respectively.
Accordingly, $\mathcal{L}=\prod_j \mathcal{L}_j$ is maximized.
Here only the {\tt BLLz} model is applied for the GLF and the obtained parameters are summarized in Table \ref{lf2}.
The parameters obtained for $\Phi(L_\gamma,z)$ are almost the same as those for the 1--3 GeV band fitting.
The mean and variance of the photon index are 2.26$\pm$0.02 and 0.26$\pm$0.04, respectively, and similar to values 2.1 and 0.26, respectively, for BL Lacs \citep{Ajel14}. 
Since the $L_\gamma$-dependence of the SED is seen in Figure \ref{modelcurve} left, the gamma-ray photon index could depend on $L_\gamma$.
Thus, the model, which has a mean photon index $\mu(L_\gamma)=\mu^{\star}+\beta\left(\log_{10}L_\gamma-43\right)$, was also tried and the result is summarized in Table \ref{lf2}.
Indeed, the $L_\gamma$-dependent coefficient $\beta=0.07\pm0.01$; since it is positive it indicates a softer GeV gamma-ray spectrum for galaxies with a higher $L_\gamma$.
This could not be due to the selection bias of {\it Fermi}/LAT detection, since there is no strong bias on the distribution of photon indices for newly detected faint AGNs in the latest 4th catalog \citep{4LAC}.

\subsection{Based on different classes of galaxies}

In the above analysis, we include {\tt RDG} and {\tt AGN} classes in the 4FGL-DR2.
We checked how the result changes if we use only {\tt RDG}, or
include {\tt CSS} and/or {\tt SSRQ}, using the {\tt BLLz} model.
The result is that the GLF parameters are not strongly affected by whether or not we include these galaxies; the normalization changed with the number of sample galaxies. 
When we include {\tt CSS} and {\tt SSRQ}, the luminosity-dependence and redshift-dependence became a little bit flatter because {\tt CSS} and {\tt SSRQ} consist of galaxies with high redshifts and luminosities.
If we exclude {\tt AGN}, these become steeper because {\tt AGN} tends to have higher redshifts and luminosities.
But the overall GLF shape does not change much.

\section{Contribution to the Extragalactic Gamma-ray Background}
\label{sec:cgb}

We calculate the contribution of gamma-ray emitting radio galaxies to the extragalactic gamma-ray background (EGB) flux using the obtained GLF as
\begin{eqnarray}\nonumber
 F_{\rm EGB}(E_0) &=& \int_{z_{\rm min}}^{z_{\rm max}}
  \frac{d^2V}{dzd\Omega} dz
  \int_{L_{\gamma, {\rm min}}}^{L_{\gamma, {\rm max}}} dL_\gamma\\
  &\times&\Phi(L_\gamma,z)  F(z,L_\gamma,E_0),
\label{CXB}
\end{eqnarray}
where $F(z,L_\gamma,E_0)$ is a flux at an energy $E_0$ in the observer frame for a source with a redshift $z$ and a gamma-ray luminosity $L_\gamma$. We apply two different methods for the calculation of $F(z,L_\gamma,E_0)$. The first method treats $F(z,L_\gamma,E_0)$ as a summation of a single power-law with various photon indices $\Gamma$ following a Gaussian distribution $G(\Gamma)$, derived in \S~\ref{EneLF}. We add a high-energy spectral cut-off at 500~GeV, since PKS~0625-354, one of the hardest $\Gamma$ radio galaxies, has a cut-off around this energy \citep{HESS18}. We use the GLF obtained in the 1--3 GeV band and extrapolate the contribution of this band to other energy bands. In the second method we treat $F(z,L_\gamma,E_0)$ as a single power-law with an average photon index in the {\it j}-th energy band and calculate the contribution in the {\it j}-th energy band using the obtained GLF in each energy band. In both cases, absorption by the extragalactic background light is taken into account by using the formula of \citet{Inou13}. 

\begin{figure}[t]
\begin{tabular}{cc}
\begin{minipage}{0.5\hsize}
\begin{center}
\includegraphics[width=8cm]{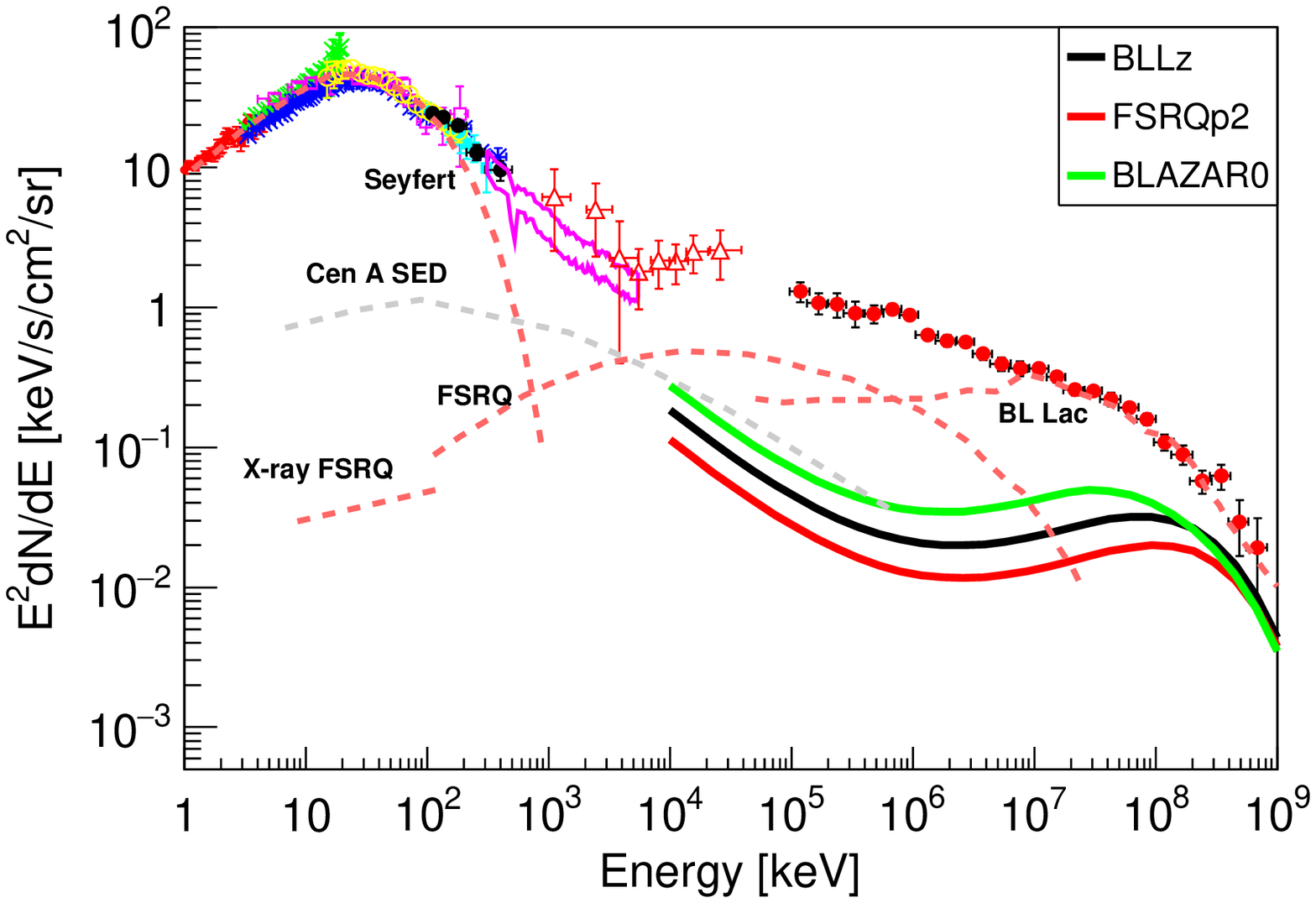}
\end{center}
\end{minipage}
\begin{minipage}{0.5\hsize}
\begin{center}
\includegraphics[width=8cm]{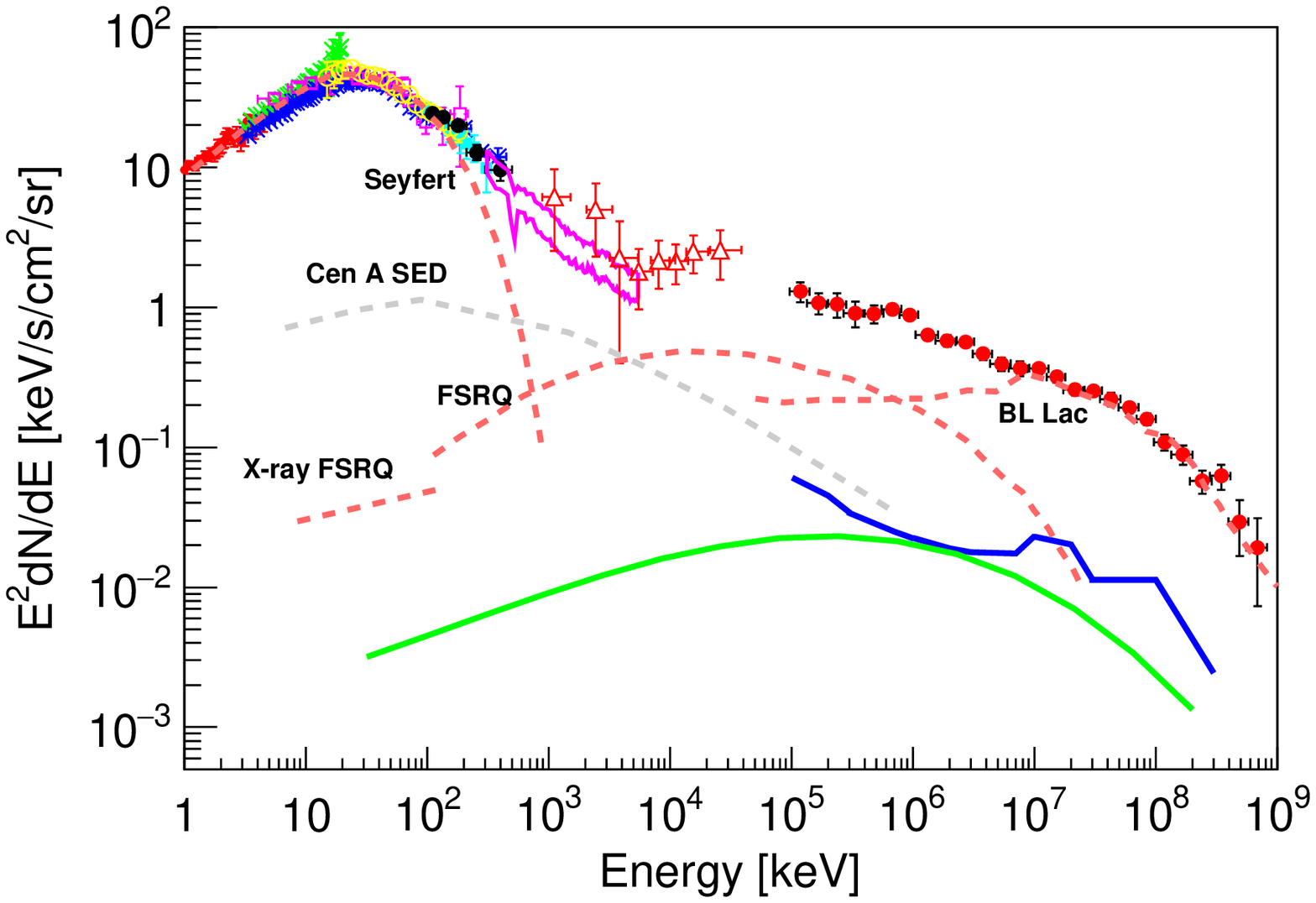}
\vspace*{0.5cm}
\end{center}
\end{minipage}
\end{tabular}
\vspace*{0.3cm}
\caption{Contribution of radio galaxies to the cosmic X-ray and gamma-ray background radiation, estimated by the best-fit LDDE models.
{\it Left}: Estimation based on a powerlaw SED whose photon index follows a Gaussian distribution. LF is using best-fit LFs ({\tt BLLz}, {\tt FSRQp2}, and {\tt BLAZAR0}) models determined in 1--3 GeV band. {\it Right}: Blue line is estimated in each of the 6 band by using {\tt BLLz} LF model determined in each band. The light green line is estimated by using a SED of our sample radio galaxies from X-ray to GeV gamma-ray band. {\tt BLLz} LF model determined in 1--3 GeV is used. {\it Both panels}: Red dashed curves denoted as Seyfert, X-ray FSRQ, FSRQ, and BL Lac are contributions of AGN to the EGB in \citet{Gill07}, \citet{Toda20}, \citet{Ajel12}, and \citet{Qu19}, respectively, Grey dashed line denoted as Cen A is a SED shape of Cen A \citep{Abdo10e}. Observational extragalactic background data are taken from \citet{Toda20} and references are therein.}
\label{egb}
\end{figure}

The left panel of Figure~\ref{egb} shows the EGB spectrum using the former method. The contribution of radio galaxies to the EGB is predicted to $\sim$ a few percent from the {\tt BLLz} LF model. This is smaller than that reported by previous studies, where radio luminosity functions were used \citep{Inou11,DiMa14}. At higher energy, the predicted spectrum could exceed the observed EGB if there is no spectral cut-off. The {\tt FSRQp2}  LF model gives a smaller contribution than {\tt BLLz}, because it predicts a smaller number of radio galaxies at high redshifts (see Figure \ref{lfcmp}). The {\tt BLAZAR0} LF model predicts a higher contribution, around 5--10\% of the EGB but still smaller than previous estimates. This is caused by a higher redshift peak in these LFs than {\tt BLLz} and {\tt FSRQp2}; this LF predicts a larger number of radio galaxies at higher redshift.

The right panel of  Figure~\ref{egb} shows the contribution based on the GLF obtained in each of the 6 energy bands by using the {\tt BLLz} model. It is almost consistent with the prediction by using a summation of a cut-off power-law with a Gaussian distribution of photon indices. It predicts a smaller contribution in the lower energy band. This may be due to the result of the assumption of the Gaussian distribution for $\Gamma$ in the left panel.

Next, by using the obtained SED shape of our sample radio galaxies from X-ray to GeV gamma-ray band in \S\ref{ana:sed}, we calculate a contribution to the EGB in the MeV gamma-ray band. Our sample radio galaxies are divided into the four luminosity groups; $10^{40-42}$, $10^{42-43}$, $10^{43-44}$, $10^{44-46}~{\rm erg\ s^{-1}}$ in the 1--3 GeV band, and an average SED shape is obtained in each luminosity range. These average SEDs are shown as light green lines in Figure \ref{modelcurve} left. Extrapolating these 4 SED shapes to arbitrary luminosity, we calculated a contribution to the EGB in the MeV energy band. The result by using the {\tt BLLz} LF model is shown in the right panel of Figure~\ref{egb}. The contribution of gamma-ray emitting radio galaxies in the MeV band is $\approx$ a few percent.

\section{Discussion}
\label{sec:dis}

\subsection{Comparison with blazars}
\label{sec:lf}

In this study, we adopt eight LDDE GLF models to fit the data. Among them, the {\tt BLLz} and {\tt FSRQp2} models reproduced the observed redshift, luminosity, and source count distributions best. These two models have the same redshift dependence as those of GeV gamma-ray BL Lacs and FSRQs, respectively, except for the peak redshift $z_c$ for the former and $p_2$ for the latter. For {\tt BLLz}, $z_c<0.1$, which is lower than the typical peak redshift of blazars and radio-quiet AGNs, where $z_c\approx2$. For{\tt FSRQp2}, $p_2>0$. Therefore, the GeV gamma-ray luminosity function of radio galaxies is similar to that of blazars, but a lower peak redshift or a negative evolution is preferred. 

\begin{figure}[t]
\begin{tabular}{cc}
\begin{minipage}{0.5\hsize}
\begin{center}
\includegraphics[width=8cm]{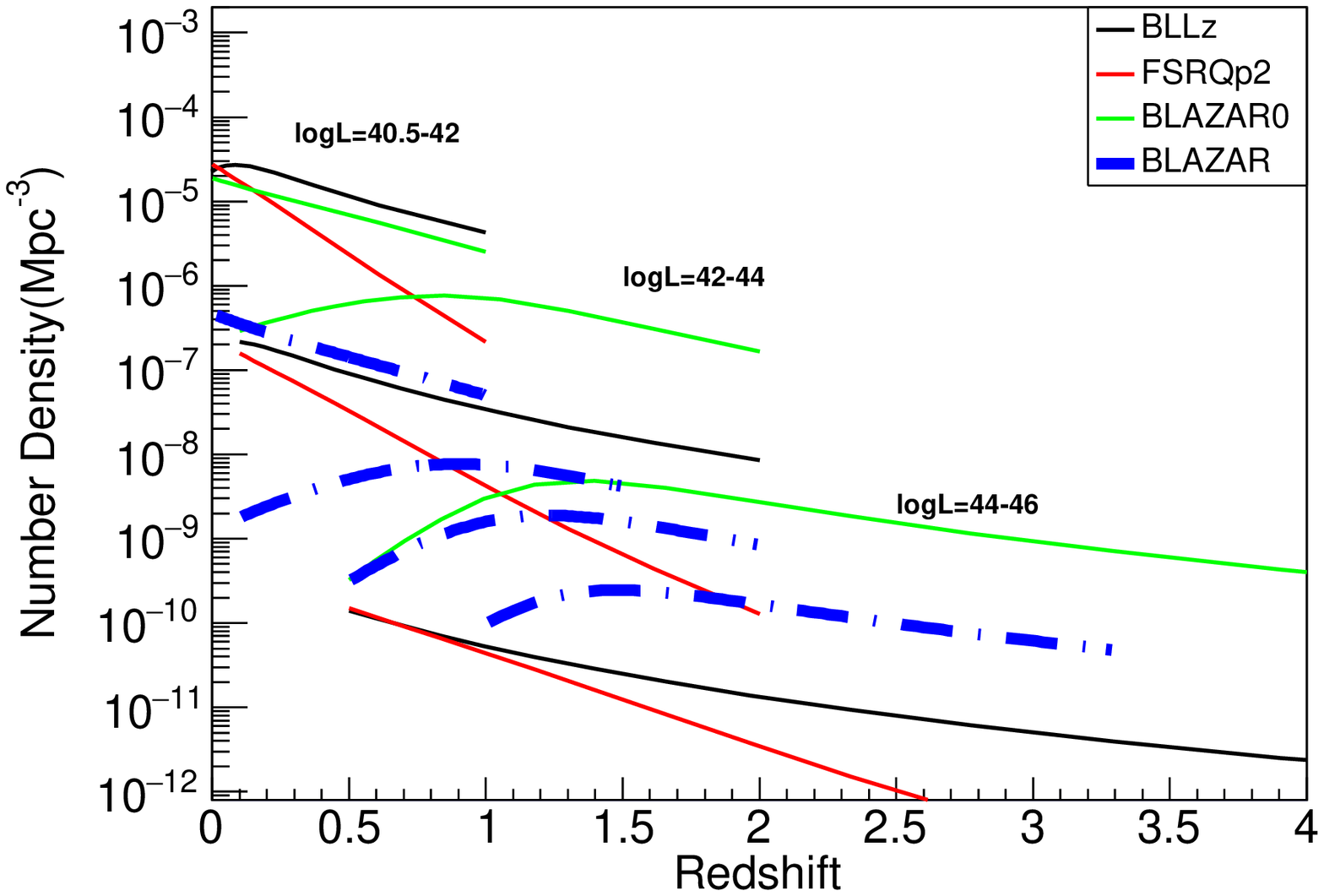}
\end{center}
\end{minipage}
\begin{minipage}{0.5\hsize}
\begin{center}
\includegraphics[width=8cm]{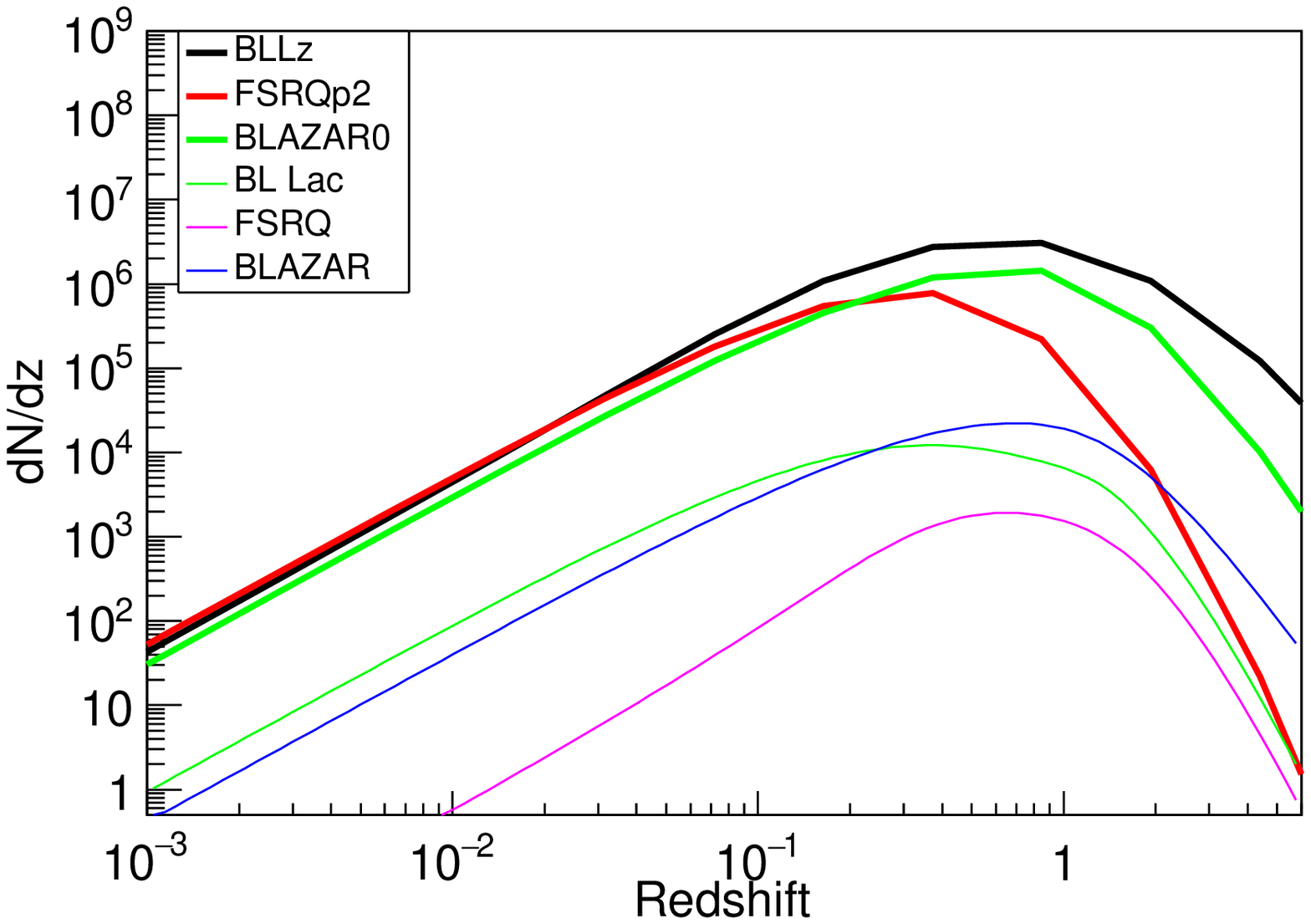}
\vspace*{0.5cm}
\end{center}
\end{minipage}
\end{tabular}
\vspace*{0.3cm}
\caption{{\it Left}: Comparison of best-fit gamma-ray lumnosity functions (as comoving number density). Colors of best-fit 3 models are the same as figure \ref{zldist}. Curves in three luminosity ranges of $10^{40.5-42}$, $10^{42-44}$, and $10^{44-46}$ erg s$^{-1}$ are plotted for each model. As a comparison, the gamma-ray luminosity function of all blazar \citep{Ajel15} is also shown as blue thick dashed lines. Four curves for all blazar correspond to luminosity ranges of $10^{43.8-46.8-44.8}$,$10^{45.8-46.9}$, $10^{46.9-47.9}$, and $10^{47.9-49.4}$ erg s$^{-1}$. 
{\it Right}: Number distribution of radio galaxies against redshift, predicted by best-fit GLF models. Model curves are calculated without sky coverage function. Colors are the same as those of left panel.}
\label{lfcmp}
\end{figure}

The left panel of Figure~\ref{lfcmp} shows a comparison of the best-fit GLFs ( {\tt BLLz} and {\tt FSRQp2}), together with those of blazars. Although these two models give similar number densities in the redshift ranges of our sample galaxies, their behavior at higher redshifts is different. 

Radio galaxies are believed to be the parent population of blazars, so it is natural to expect that their LF shape is similar. Our sample is dominated by low power radio galaxies with similar radio power to that of FR-I galaxies, the parent population of BL Lacs. However, as seen in the left panel of Figure~\ref{lfcmp}, a clear difference in LF shape exists between gamma-ray emitting radio galaxies and blazars. Compared to the blazar GLF, both {\tt BLLz} and {\tt FSRQp2} models give smaller number counts at higher redshifts, and larger number counts at lower redshifts. We also show the model curve of  {\tt BLAZAR0}, which follows the GLF of the combined population (BL Lac and FSRQ) of gamma-ray blazars \citep{Ajel15}. Even with this model, the GLF curves of radio galaxies appear to be different from that of blazars.

Low-luminosity high-synchrotron-peaked BL Lacs (HSPs) show a strong negative evolution \citep{Ajel14} like gamma-ray emitting radio galaxies. If HSPs are the parent population of gamma-ray emitting radio galaxies, this may be a possible reason for their GLF shapes. However, HSPs are a specific population of blazars. In fact, low-z gamma-ray emitting radio galaxies do not necessarily have a high peak frequency or a hard GeV gamma-ray photon index as seen in the right side of Figure \ref{zlph}. 

Another possible interpretation is that emission regions of gamma-ray radio galaxies seen by {\it Fermi} may be different from those of blazars. Then, an additional component appears in low redshift GLFs. Here, recent gamma-ray observations found evidence that a large population of low-z low-luminosity radio galaxies might be due to gamma-ray emission from sources other than the core jet. Cen~A and Fornax~A show extended gamma-ray emission from their radio lobes \citep{Abdo10d, Acke16}. Cen~A also shows the existence of kpc-scale extended gamma-ray emission \citep{Prok19,HESS20}, where various models are proposed to explain the origin \cite[e.g.,][]{Tana19, Sudoh2020ApJ, Rieger2021ApJ}. Hot spot emission could also contribute to gamma-ray emission in radio galaxies, as suggested for M87 HST-1 \citep{Harr09,Imaz21}. This non-core emission could make up a significant fraction of GeV detected gamma-rays from radio galaxies. Therefore, the difference in GLF shapes between radio galaxies and blazars could be due to an additional component from non-core emission, appearing in low redshifts.

To investigate the relation between blazars and gamma-ray emitting radio galaxies further, in the right panel of Figure \ref{lfcmp}, we show the redshift distribution of radio galaxies, obtained by using Equation \ref{dNdzdNdL} but without sky coverage. Thus, non-detected galaxies are considered.
We also show the redshift distribution of blazars for comparison. The number of radio galaxies is 10--30 times larger than that of BL Lacs.

From the viewpoint of population, the comparison with the curve of BL Lacs is natural since low power radio galaxies dominate our samples. However, the ratio of 10--30 is unexpectedly small when we consider the beaming effect. Our sample radio galaxies have a lower gamma-ray luminosity by a factor of $\sim10^{4-5}$ than blazars. Considering the beaming correction factor, $\delta^4$ where $\delta$ is a beaming factor, radio galaxies in 4FGL-DR2 should have a smaller $\delta$ by a factor of $\sim10$ than blazars. In that case, the ratio should be $\sim100$. \citet{Inou11} proposed that only radio galaxies with a viewing angle of $\la24^{\circ}$ are seen in the GeV gamma-ray band.

Alternatively, if {\it Fermi}/LAT misses faint radio galaxies at the lowest luminosity end, the number ratio of GeV emitting radio galaxies to BL Lacs could be underestimated. Even for nearby galaxies, the detection limit in luminosity is around $10^{41}$ erg s$^{-1}$, and thus radio galaxies with a luminosity of $10^{40-41}$ erg s$^{-1}$ could be a hidden population.

\subsection{Comparison with radio galaxies seen in the radio band}
\label{sec:pop}

The {\tt RLF} model, which is obtained in the radio band, cannot reproduce the observed  distributions of gamma-ray emitting radio galaxies at all (See Figure \ref{zldist}). This indicates that a population of GeV emitting radio galaxies is different from or represents a peculiar population of radio galaxies seen in the radio band. This section investigates the differences between gamma-ray detected and non-detected radio galaxies.

\citet{Ming14} compiled X-ray properties of the 2 Jy complete sample of radio galaxies, where only 4 out of 46 galaxies are GeV gamma-ray emitting ones. Some of our sample radio galaxies have fainter X-ray fluxes comparing to the \citet{Ming14} sample. Thus the smaller fraction of our sample compared with the \citet{Ming14} sample is not due to the X-ray flux limit, but because only a small fraction of radio galaxies are gamma-ray loud. Gamma-ray quiet radio galaxies do not follow the correlation between X-ray and gamma-ray luminosity shown in Figure~\ref{lglx}.
This ratio $(4/46)$ is similar to the population ratio estimate in \citet{Inou11}, which estimated the ratio as $0.088$ based on the comparison of the number counts between the radio and GeV gamma-ray bands. They claimed that a large population of radio galaxies are faint in the gamma-ray band due to a large viewing angle, which makes the jet faint. As seen in Figure \ref{lglr}, the gamma-ray luminosity is different by more than two orders of magnitude among galaxies with the same radio luminosity, when considering those not detected by {\it Fermi}/LAT. Such a large difference of gamma-ray luminosity could be caused by the beaming effect; gamma-ray emission comes from the jet core and thus the gamma-ray luminosity is strongly affected by the beaming effect, while this radio emission comes not only from the core but also an extended outer region and thus it is less beamed.

GeV emitting radio galaxies of our sample comprise 22 FR-Is and 14 FR-IIs. 
Unclassified galaxies in our sample do not tend to have a higher radio power like FR-II.
Therefore, observationally low radio power galaxies like FR-I are the dominant class in the GeV gamma-ray band. This is contrary to the radio-band population; there are 6 FR-I and 33 FR-II in the 2 Jy complete sample of radio galaxies \citep{Dick08}. Therefore, it is clear that the population of gamma-ray emitting radio galaxies lacks high radio power galaxies like FR-IIs.

In contrast, there are 8 CSS and SSRQ galaxies in the GeV gamma-ray band, while there are 6 CSS galaxies in the 2 Jy radio sample. Therefore, the fraction of CSS and SSRQ is similar in the radio and GeV gamma-ray bands. CSS and SSRQ are high-luminosity radio galaxies, and thus an intermediate class between FR-II and FSRQ.

\begin{figure}[t]
\begin{tabular}{cc}
\begin{minipage}{0.5\hsize}
\begin{center}
\includegraphics[width=8cm]{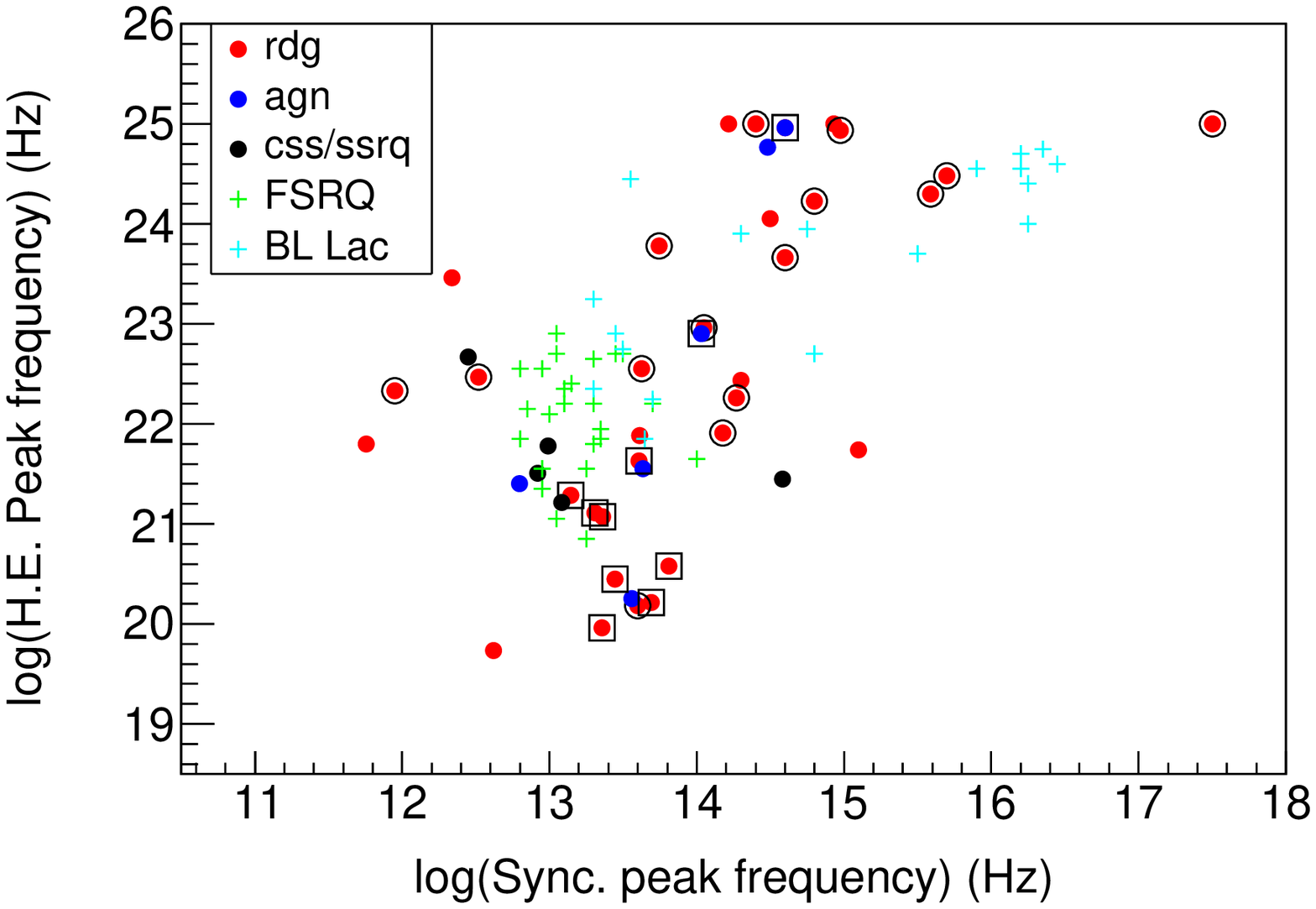}
\end{center}
\end{minipage}
\begin{minipage}{0.5\hsize}
\begin{center}
\includegraphics[width=8cm]{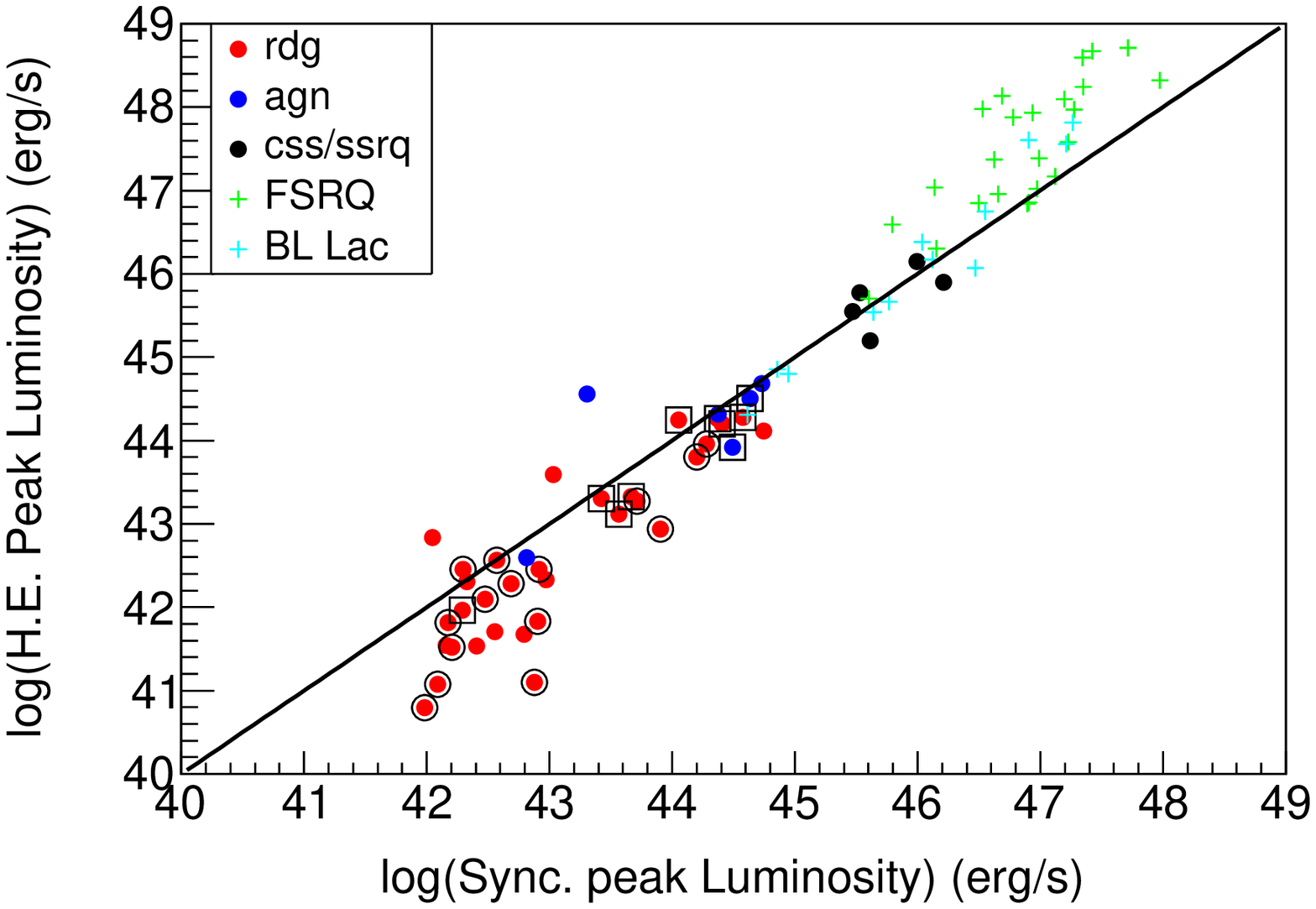}
\vspace*{0.5cm}
\end{center}
\end{minipage}\\
\begin{minipage}{0.5\hsize}
\begin{center}
\includegraphics[width=8cm]{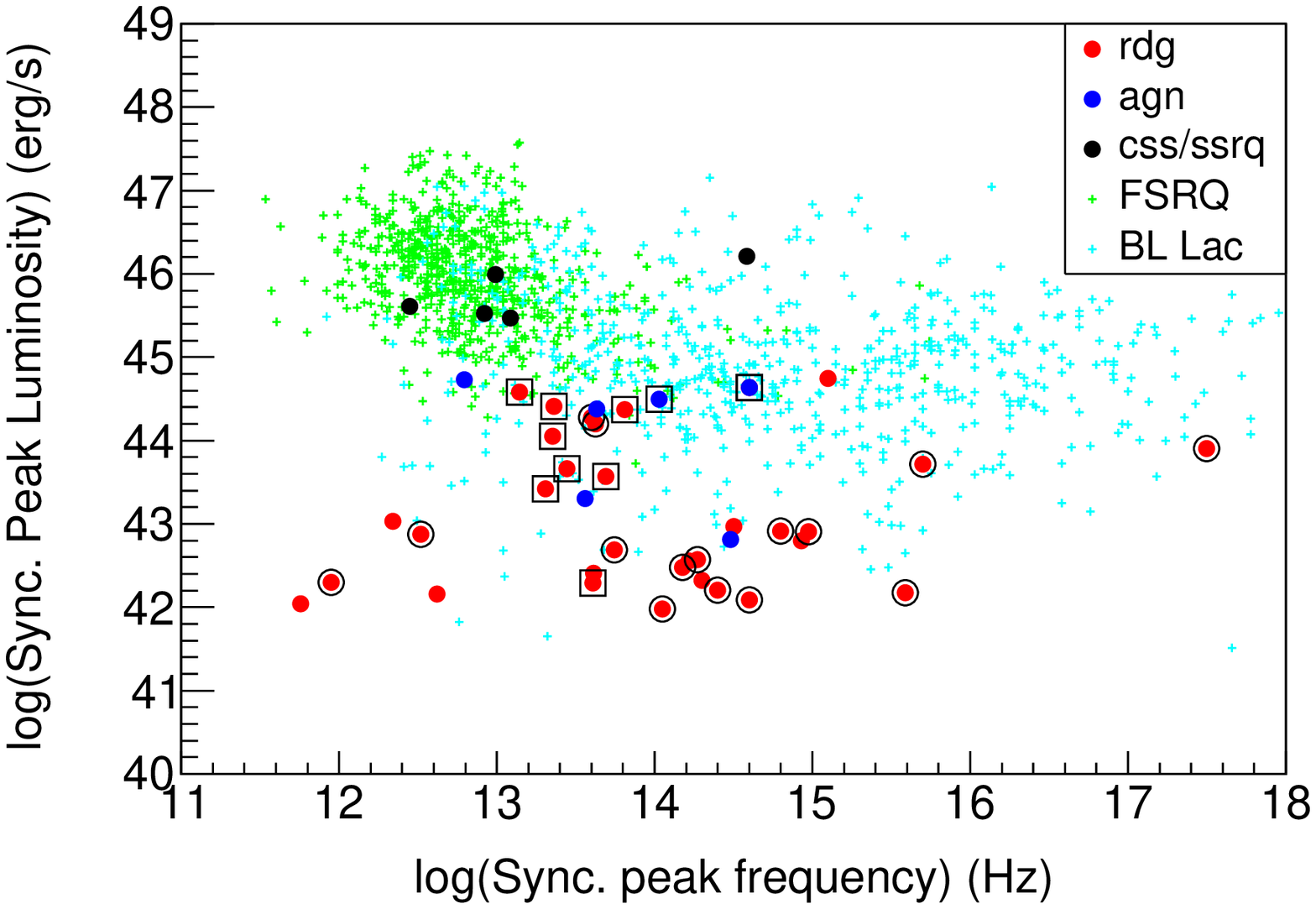}
\end{center}
\end{minipage}
\begin{minipage}{0.5\hsize}
\begin{center}
\includegraphics[width=8cm]{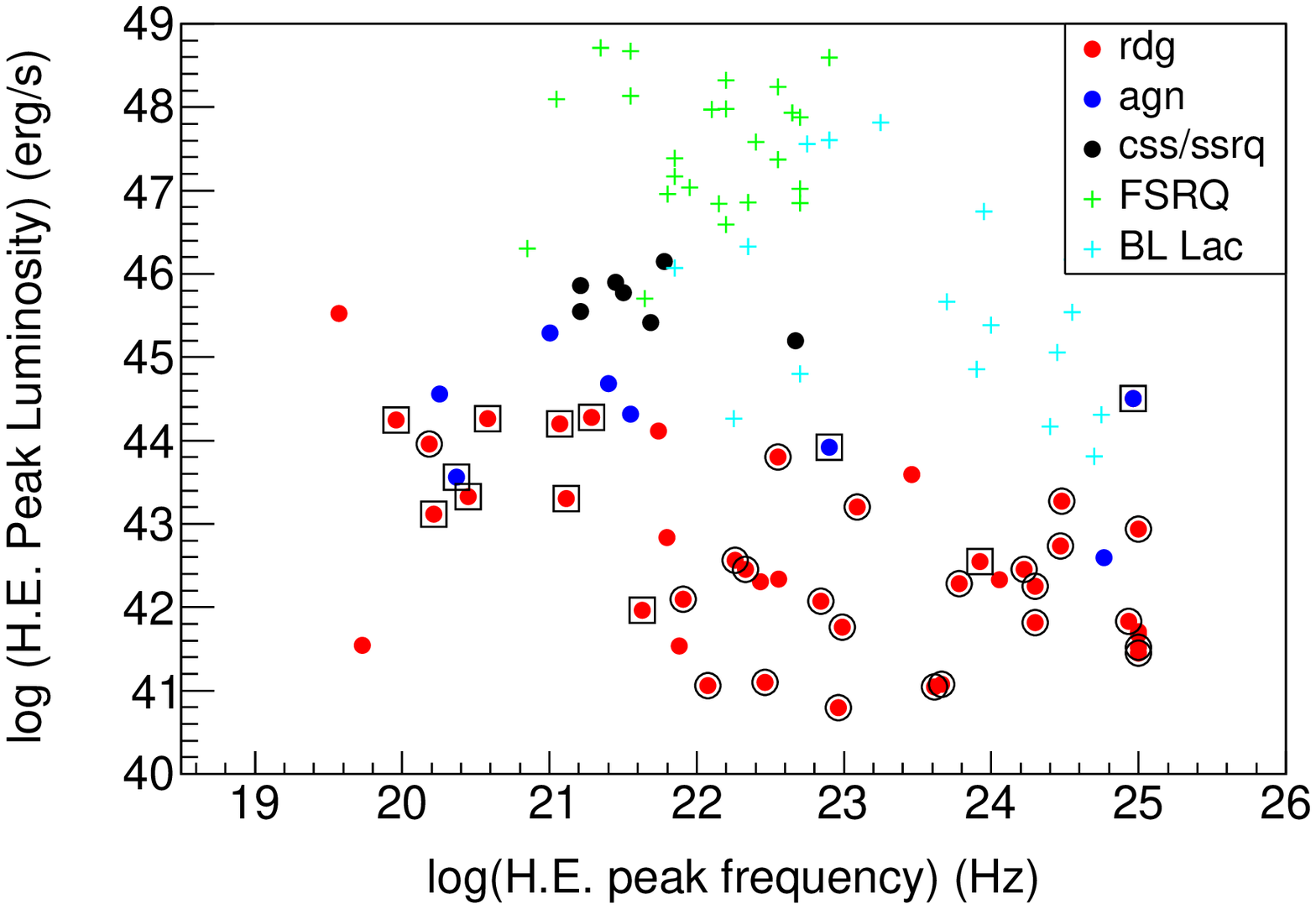}
\vspace*{0.5cm}
\end{center}
\end{minipage}
\end{tabular}
\vspace*{0.3cm}
\caption{{\it Top-Left}: Synchrotron peak frequency vs high energy component (H.E.) peak frequency. {\it Top-right}: Synchrotron peak luminosity vs high energy component peak luminosity.  {\it Bottom-left}: Peak frequency vs peak luminosity for Synchrotron component. {\it Bottom-right}: Peak frequency vs peak luminosity for high energy component. 
Red, blue, and black circles represent {\tt rdg}, {\tt agn}, and {\tt css/ssrq}, respectively. Data enclosed by circles or squares represent FR-I or FR-II, respectively. 
Thin green and thin blue crosses represent FSRQ and BL Lac, respectively.}
\label{sedpeak}
\end{figure}

\begin{figure}[t]
\begin{tabular}{cc}
\begin{minipage}{0.5\hsize}
\begin{center}
\includegraphics[width=8cm]{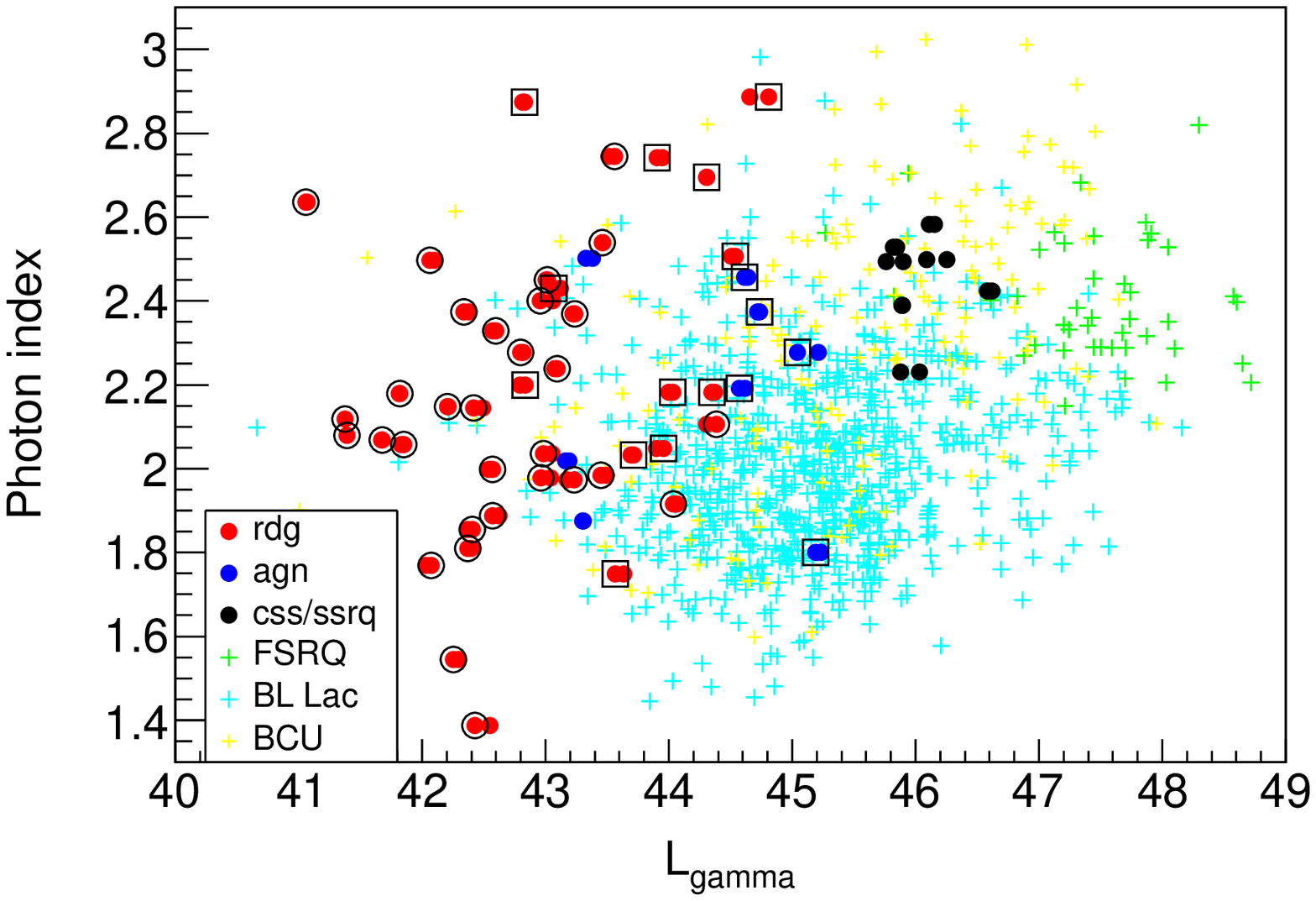}
\end{center}
\end{minipage}
\begin{minipage}{0.5\hsize}
\begin{center}
\includegraphics[width=8cm]{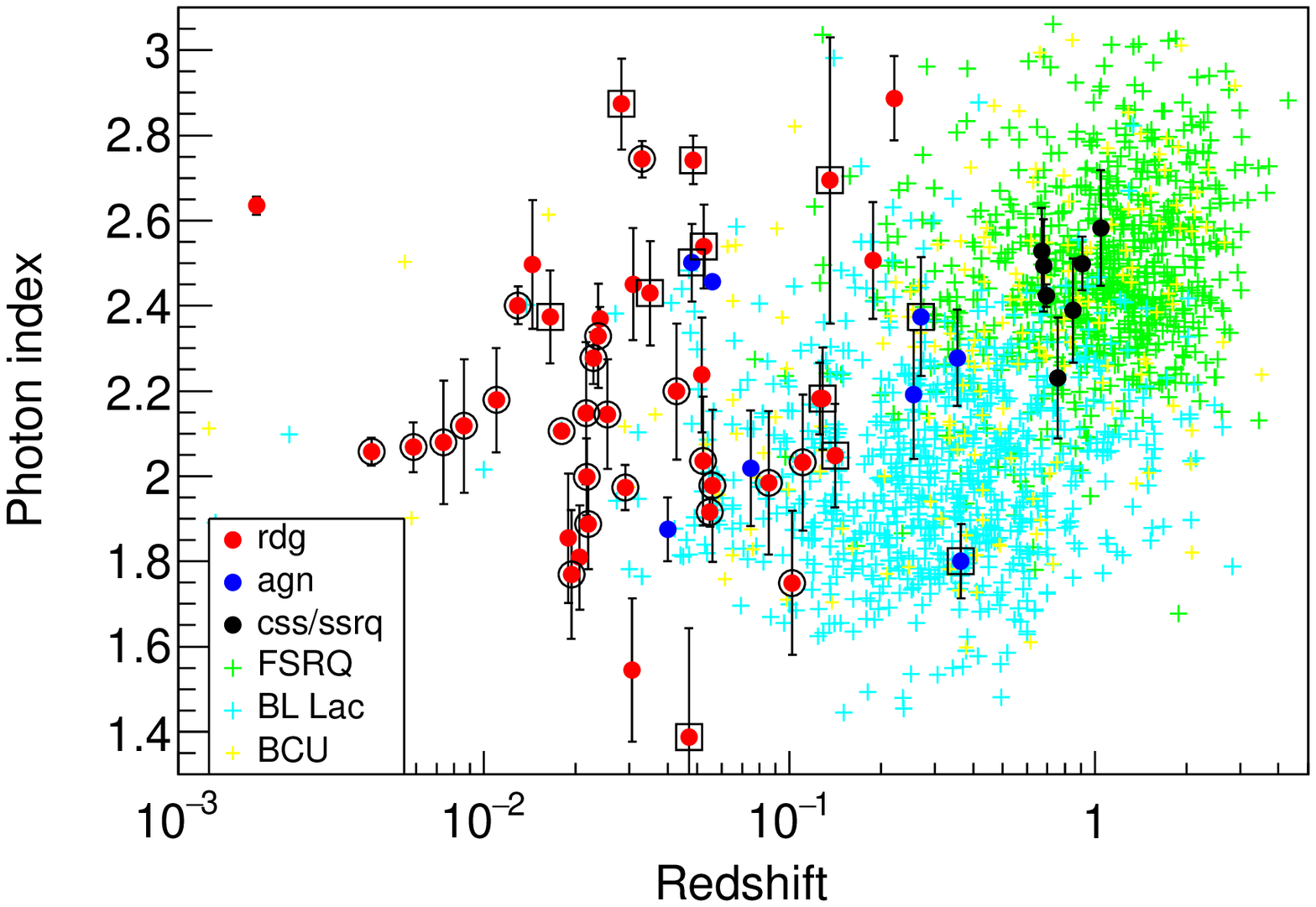}
\vspace*{0.5cm}
\end{center}
\end{minipage}
\end{tabular}
\vspace*{0.3cm}
\caption{{\it Left}: Gamma-ray luminosity vs Gamma-ray photon index for 4FGL-DR2 MAGNs, together with blazars. Colors and symbols are the same as those of figure \ref{sedpeak}.}, except for the yellow crosses which are the BCUs. {\it Right}: Redshift vs Gamma-ray photon index for 4FGL-DR2 MAGNs, together with blazars. Colors and symbols are the same as those of the left panel.
\label{zlph}
\end{figure}

As seen in Figure \ref{zlph}, all CSS and many FR-II galaxies show steep gamma-ray spectra. Many galaxies having steep spectra are FR-II. Therefore, it is likely that gamma-ray SED of FR~IIs tends to drop down below the {\it Fermi}/LAT energy band due to the low peak frequency of their high energy component. This is also seen in Figure \ref{sedpeak}, where FR-IIs tend to have lower peak frequencies.

It has been suggested that many FSRQs are not detected by {\it Fermi}/LAT due to their low peak frequency of the IC component \citep{Pali17}. If FR-IIs are a parent population of FSRQs, and the IC peak appears lower due to less beaming, they are more likely to be missed in the GeV band. Note that 80\% of radio galaxies are FR-II in an X-ray flux-limited {\it Swift}/BAT sample \citep{Rune20}. This is similar to the case of a radio flux-limited sample \citep{Dick08}; X-ray surveys do not have a strong bias for radio galaxies. In the X-ray band, FR-IIs have high luminosity disk/corona emission like 3C 111 and 3C 120 \citep{Fuka15}.  Such objects have jet emission that extends up to at least the GeV gamma-ray band. Therefore, if many FR-IIs will be detected in the MeV gamma-ray band by future missions such as COSI \citep{COSI22}, AMEGO-X \citep{Henr22}, and GRAMS \citep{Aramaki2020APh}, the picture of the population will be more established from the viewpoint of the SED shape.

The other effect to make FR-IIs fainter in the GeV gamma-ray band might be the jet structure. FR-Is tend to have a structured jet, and thus emission from a sheath region with a lower Lorentz factor is less beamed, and thus GeV gamma-rays can be observed even if the jet is misaligned \citep{Ghisellini2005}. FR-II emission is more beamed and thus the misaligned situation makes it less luminous. \citet{Keen20} discussed a similar model to explain the synchrotron peak frequency-luminosity relation of blazars and radio galaxies.

Figure \ref{sedpeak} shows that the peak luminosity of the X-ray to gamma-ray SED is not significantly higher than that of the synchrotron peak luminosity for any radio galaxy, unlike in FSRQs. The external Compton component has a different beaming pattern than synchrotron or synchrotron Self-Compton \citep{Derm95}. The Compton dominance goes as $\max\left(\delta^2 u_{\rm ext}, u_{\rm sy}\right) / u_{\rm B}$, where $\delta$ is the Doppler factor, $u_{\rm ext}$, $u_{\rm sy}$, and $u_{\rm B}$ are the energy densities of the external radiation field, synchrotron photons, and magnetic field, respectively \citep{Fink13}. For radio galaxies with high viewing angles, $\delta$ will be small compared to blazars. Therefore, it might be that the external Compton component cannot be observed for FR-IIs because it is below the synchrotron self Compton component.

\citet{Keen20} proposed that FSRQs, LBLs, and high-excitation radio galaxies (HERG; radio galaxies with high-excitation optical emission lines) have a strong jet with a wide range of jet power, while intermediate- or high-frequency-peaked BL Lacs (IBL or HBL) and low-excitation radio galaxies (LERG; radio galaxies with low-excitation optical emission lines) have weak jets with low powers. The FSRQs, LBLs, and HERGs could correspond to FR-IIs, and the IBLs, HBLs, and LERGs to FR-Is. However, some of gamma-ray emitting radio galaxies (PKS 0625-354, 3C 264, and TXS 1516+064) have high peak synchrotron frequencies of $10^{15.5-17.5}$ Hz, and their high-energy components have peak frequencies of $10^{24.5-25.0}$ Hz. Others have similar high peak frequencies of the high-energy component. These galaxies have a low gamma-ray luminosity compared to BL Lacs, and thus they do not follow the model of \citet{Keen20}.

\subsection{The {\tt AGN} class in 4FGL-DR2}

The 4FGL-DR2 sample contains the {\tt AGN} class, whose properties and counterparts are not well studied in other wavelengths. These galaxies cannot be classified as blazars with uncertain type ({\tt BCU} class in 4FGL-DR2), since they do not satisfy the criteria of blazars. Their SEDs are similar to that of compact radio sources, but the uncertainty is large. As shown in Figure \ref{zlph}, {\tt AGN} seem to be divided into low luminosity ones and high luminosity ones, corresponding to FR-I and FR-II, respectively.

One candidate population of this class is FR-0s \citep{Ghis11}, where the radio emission is compact. In fact, NVSS radio images of most of these galaxies on the {\tt SkyView}\footnote{https://skyview.gsfc.nasa.gov/current/cgi/titlepage.pl} do not exhibit a clear jet structure.
\citet{Torr18} systematically studied X-ray properties of FR-0s, and found that their X-ray properties are almost the same as those of FR-Is. Therefore, some low luminosity {\tt AGN} could be FR-0s. \citet{Pali21} reported a possible gamma-ray emitting population of FR-0s in a stacking analysis of {\it Fermi}/LAT data.
\citet{Itoh20} reported that there is a significant number of elliptical-like blazars in their blazar catalog. These objects are low-luminosity BL Lacs and apparently look like elliptical galaxies. Therefore, such objects with misaligned jet are one candidate population of the {\tt AGN} class. 

\citet{Damm15} reported that PKS 0521-36, the brightest among the {\tt AGN} class objects, has properties intermediate between broad-line radio galaxies (BLRGs) and SSRQs. BLRGs have a high accretion rate and thus higher luminosity. Therefore, high-luminosity {\tt AGN} could be a transition class between FR-IIs and blazars.

\subsection{X-ray to gamma-ray spectra}
\label{sec:sed}

The study of X-ray emission from radio galaxies is important for constraining the broad-band emission from jets, because so far jet emission from radio galaxies has been primarily detected in the radio and gamma-ray bands. \citet{Fuka15} systematically analyzed {\it Suzaku} X-ray spectra of 1FGL radio galaxies. They reported that the X-ray emission is dominated by disk/corona emission for HERGs and jet emission for LERGs.

Since HERGs have higher accretion rates, their luminosities should be higher. HERGs tend to have harder X-ray spectra, while LERGs tend to have softer spectra. This is consistent with the result of high energy component SEDs from the X-ray to GeV gamma-ray band (Figure \ref{sedpeak}). Higher luminosity radio galaxies show lower peak frequencies for the high energy component, where the X-ray emission is the low-energy tail of the inverse Compton scattering component from the jet and/or disk/corona emission. Bright disk/corona X-ray emission can make peak frequencies of a high energy component lower, and thus, like the blazar sequence, could be enhanced for high energy components of radio galaxies. 

LERGs show a higher peak frequency for their high energy components, where the X-ray emission is the high-energy tail of jet synchrotron emission. A detailed systematic X-ray study with a classification of radio galaxies will be presented by a separated paper.

The Gamma-ray spectral index has a wide distribution, from $1.5$ to $3.0$. This range is similar to that of blazars, and it is caused by a variety of high energy component peak frequencies. This seems to be consistent with the view that radio galaxies are the parent population of blazars. However, as we discussed in \S~\ref{sec:lf}, the GLF shapes are different. This indicates that some of gamma-ray emitting radio galaxies are not just misaligned blazars.

For galaxies with hard GeV gamma-ray spectra, cut-offs are not seen in the GeV band. The gamma-ray brightest galaxy with a hard spectrum is PKS 0625-354, and its cut-off is detected around 0.5 TeV by HESS \citep{HESS18}. If the cut-off energy was much higher, the contribution to the EGB could exceed the observed flux, as shown in Figure \ref{egb}. This spectral cut-off is also similar to that of blazars, and could be due to the Klein-Nishina effect.

The gamma-ray spectrum of Cen~A has an upturn between the GeV and TeV \citep{HESS18b}, suggesting two emission components. In fact, TeV gamma-ray emission from the Cen A kpc-scale lobe was detected by HESS \citep{HESS20}, suggesting particle acceleration in the kpc-scale diffuse or knot regions. \citet{Rult20} surveyed the {\it Fermi}/LAT spectra of radio galaxies, but only Cen~A shows an upturned spectrum.  Considering that the peak frequency of the high energy component of Cen~A is the lowest among radio galaxies and thus it is the faintest in gamma-rays, future surveys may find other gamma-ray faint radio galaxies with upturned spectra.

\subsection{Contribution to EGB}
\label{sec:egb}

Various studies estimated the contribution of radio galaxies to the EGB \citep{Inou11, DiMa14, Hooper2016JCAP, Stecker2019ApJ, Blanco2021arXiv}. \citet{Inou11} estimated this contribution by using radio galaxies in the 1FGL, 3FGL. FL8Y (a precursor to 4FGL), and 4FGL. They estimated the contribution in a similar way to what \citet{Inou11} did, by assuming that the GLF shape is the same as that determined by radio observation of radio galaxies \citep[that is, {\tt RLF;}][]{Will01}, and that the gamma-ray luminosity correlates with radio luminosity. We determined the GLF normalization by using the number of gamma-ray emitting radio galaxies, and the contribution to the EGB was estimated to be several to several tens of percent, dependent on the gamma-ray spectral index.

Our estimation is based on only gamma-ray results and has as few assumptions as possible. Although some GLF parameters are fixed, we fitted the distributions of gamma-ray luminosities and redshifts by 8 model GLFs, and confirmed that they reproduce the data. In addition, we considered an energy dependence of the cEGB ontribution both by using a Gaussian distribution for the gamma-ray photon index, and by estimating the contribution in each of 6 gamma-ray energy bands independently. As a result, the contribution is estimated to be 1--10\% of the 0.1--300 GeV background, dependent on the GLF shape, especially for the redshift dependence of the GLF. In other words, the radio galaxy contribution to the EGB is dominated by unresolved (non-detected) radio galaxies. In order to constrain the dark matter contribution to the EGB, it is important that the contribution of radio galaxies is estimated as accurately as possible.

Furthermore, by obtaining the X-ray to GeV gamma-ray SED, we estimated the contribution of GeV gamma-ray emitting radio galaxies to the EGB in the MeV band to be $<$1\%. However, as discussed in \S\ref{sec:pop}, there could be a large number of radio galaxies hidden in the GeV gamma-ray band. Considering that only 10\% of the 2 Jy flux-limited sample are detected in the GeV band, but most of them are detected in the X-ray band, the contribution of radio galaxies to the EGB in the MeV band could be larger by a factor of up to 10.

Below 10 MeV, the EGB spectrum becomes softer with a turn over around 5 MeV, while the SEDs of GeV gamma-ray emitting radio galaxies are assumed to connect directly to the X-ray band, and thus the contribution below 1 MeV becomes much smaller than 1\%. Radio-quiet AGNs dominate below several 100~keV \citep{Ueda2003, Gill07, Ueda14} because there are many more of them than radio galaxies and their spectra show cut-offs around several 100 keV. These cut-offs are due to thermal emission from the disk/corona, while non-thermal emission from the disk/corona is suggested and could contribute to the MeV EGB \citep{Inou08,Inou19,Murase2020PhRvL}.

Considering radio galaxies make up 7--10\% of the objects in the {\it Swift}/BAT AGN catalog \citep{Pane16,Gupt18}, the contribution of radio galaxies to the cosmic X-ray background (CXB) could be about 10\% of the contribution of Seyfert galaxies \citep{Gill07}.
This is 100 times larger than the contribution predicted from GeV emitting radio galaxies.
As discussed in \S\ref{sec:pop}, numerous GeV-quiet radio galaxies would have SEDs similar to that of Cen A, as shown in Figure \ref{egb}.
If that is the case, the contribution to the CXB could be 100 times larger than our prediction curve in figure \ref{egb} right.
Note that, in such a case, the contribution of radio galaxies to the EGB becomes comparable to that of FSRQs around 1 MeV.
A recent study of contributions of FSRQs to the MeV EGB suggested that FSRQ cannot explain the MeV EGB completely \citep{Toda20}. Therefore, future MeV gamma-ray observations with such as COSI, AMEGO-X, and GRAMS could detect many FSRQs and radio galaxies to resolve this issue.

\section{Conclusions}
\label{sec:con}

Utilizing the 4FGL-DR2 catalog, we systematically studied the X-ray to GeV gamma-ray spectra of 61 gamma-ray emitting radio galaxies. We found an anti-correlation between peak frequency and peak luminosity for the high energy component of sampled radio galaxies. This anti-correlation feature is also known to exist in blazars \citep[see e.g.,][]{Fink13,Ghisellini2017}, although its origin is controversial. We note that if the disk/corona emission, rather than jet emission, dominates the X-ray fluxes of high-luminosity radio galaxies \citep{Fuka15} it would enhance this feature.

Comparing with radio and X-ray selected radio galaxy samples, the gamma-ray selected sample of radio galaxies lacks high radio luminosity galaxies like FR-II galaxies at the same radio flux threshold. This implies FR-II galaxies appear fainter in the gamma-ray band due to the beaming effect and/or softer SED shape. Future MeV gamma-ray observations will be crucial to elucidating this discrepancy.

We further explored the cosmological evolution of gamma-ray emitting radio galaxies using the same sample. For the first time, we construct the gamma-ray luminosity function of radio galaxies using gamma-ray data only. We found that gamma-ray emitting radio galaxies favor negative evolution at all luminosity ranges, similar to HSPs. However, this trend is different from all blazars, which are on-axis radio galaxies. Therefore, gamma-ray photons may originate in different regions in gamma-ray loud radio galaxies and blazars.

By combining the gamma-ray spectra and gamma-ray luminosity functions of radio galaxies, we also estimated their contribution to the extragalactic gamma-ray background radiation. The expected contribution is about 1--10\%. However, considering hidden GeV gamma-ray emitting radio galaxies, the contribution to EGB could be around 10\% in the MeV band.

\acknowledgements
Y.I.\ is supported by JSPS KAKENHI Grant Number JP18H05458 and JP19K14772.  J.F.\ is supported by NASA under contract S-15633Y.

%\begin{landscape}
%\begin{scriptsize}
%\startlongtable
\begin{deluxetable}{rllcccrcc}
\tablecaption{Our sample 4FGL-DR2 radio galaxies}
%\hspace{-1.8cm}
\tablehead{
\colhead{No.} & \colhead{4FGL~name} & \colhead{galaxy~name} & \colhead{class$^a$} & \colhead{$z$} & \colhead{${\rm log}_{10}L_{1.4}^b$} & \colhead{RG$^c$} & \colhead{flag$^d$}  
}
\startdata
 1 & J0009.7-3217 & IC 1531 & rdg & 0.025 & -0.08 & I$^{\rm B18}$ & 001110+ \\ 
 2 & J0013.6+4051 & 4C +40.01 & agn & 0.255 & 2.50 & --$^{\rm }$ & 011100+ \\ 
 3 & J0038.7-0204 & 3C 17 & rdg & 0.220 & 2.39 & I/II$^{\rm ZB95}$ & 111000+ \\ 
 4 & J0057.7+3023 & NGC 315 & rdg & 0.016 & -0.44 & II$^{\rm ICS99}$ & 011110+ \\ 
 5 & J0153.4+7114 & TXS 0149+710 & rdg & 0.022 & -0.49 & I$^{\rm L01}$ & 011111- \\ 
 6 & J0237.7+0206 & PKS 0235+017 & rdg & 0.022 & -1.07 & I$^{\rm CB88I}$ & 000110+ \\ 
 7 & J0308.4+0407 & NGC 1218 & rdg & 0.029 & 1.13 & I$^{\rm ZB95}$ & 011111+ \\ 
 8 & J0312.9+4119 & B3 0309+411B & rdg & 0.136 & 1.52 & II$^{\rm ICS99}$ & 001110- \\ 
 9 & J0316.8+4120 & IC 310 & RDG & 0.019 & -0.86 & H/T$^{\rm M93}$ & 001111- \\ 
10 & J0319.8+4130 & NGC 1275 & RDG & 0.018 & 1.20 & I$^{\rm ZB95}$ & 111111- \\ 
11 & J0322.6-3712e & Fornax A & RDG & 0.006 & -2.30 & I$^{\rm ZB95}$ & 011111+ \\ 
12 & J0334.3+3920 & 4C +39.12 & rdg & 0.021 & -0.97 & 0$^{\rm RBC20}$ & 001111- \\ 
13 & J0418.2+3807 & 3C 111 & rdg & 0.049 & 1.64 & II$^{\rm OL89}$ & 111100- \\ 
14 & J0433.0+0522 & 3C 120 & RDG & 0.033 & -0.16 & I$^{\rm OL89}$ & 111100+ \\ 
15 & J0519.6-4544 & Pictor A & rdg & 0.035 & 1.25 & II$^{\rm ZB95}$ & 111110+ \\ 
16 & J0521.2+1637 & 3C 138 & css & 0.759 & 4.34 & C$^{\rm LRL83}$ & 001110- \\ 
17 & J0522.9-3628 & PKS 0521-36 & AGN & 0.056 & 1.37 & I/II$^{\rm B16}$ & 111111+ \\ 
18 & J0627.0-3529 & PKS 0625-35 & rdg & 0.055 & 0.88 & I$^{\rm W04}$ & 111111- \\ 
19 & J0708.9+4839 & NGC 2329 & rdg & 0.019 & -0.24 & I$^{\rm D21}$ & 001110+ \\ 
20 & J0758.7+3746 & NGC 2484 & rdg & 0.043 & 1.05 & I$^{\rm G94}$ & 001100+ \\ 
21 & J0840.8+1317 & 3C 207 & ssrq & 0.681 & 3.72 & II$^{\rm LRL83}$ & 111100+ \\ 
22 & J0858.1+1405 & 3C 212 & ssrq & 1.048 & 4.17 & II$^{\rm LRL83}$ & 011000+ \\ 
23 & J0910.0+4257 & 3C 216 & css & 0.670 & 3.87 & II$^{\rm P06}$ & 111000+ \\ 
24 & J0931.9+6737 & NGC 2892 & rdg & 0.023 & -0.24 & I$^{\rm J82I}$ & 111110+ \\ 
25 & J0958.3-2656 & NGC 3078 & rdg & 0.009 & -1.29 & I$^{\rm WH84I}$ & 001100+ \\ 
26 & J1012.7+4228 & B3 1009+427 & agn & 0.365 & 1.56 & II$^{\rm K18}$ & 001111+ \\ 
27 & J1116.6+2915 & B2 1113+29 & rdg & 0.047 & 1.00 & II$^{\rm ZB95}$ & 000010+ \\ 
28 & J1118.2-0415 & PMN J1118-0413 & agn & 0.000 & -- & --$^{\rm }$ & 111100+ \\ 
29 & J1144.9+1937 & 3C 264 & rdg & 0.022 & 0.81 & I$^{\rm LRL83}$ & 001111+ \\ 
30 & J1149.0+5924 & NGC 3894 & rdg & 0.011 & -0.04 & I$^{\rm X95I}$ & 001110+ \\ 
31 & J1219.6+0550 & NGC 4261 & rdg & 0.007 & 0.31 & I$^{\rm ZB95}$ & 001110+ \\ 
32 & J1230.8+1223 & M 87 & rdg & 0.004 & -1.70 & I$^{\rm ZB95}$ & 111111+ \\ 
33 & J1236.9-7232 & PKS 1234-723 & rdg & 0.024 & 0.18 & I$^{\rm FJ02}$ & 011100- \\ 
34 & J1306.3+1113 & TXS 1303+114 & rdg & 0.086 & 0.88 & I$^{\rm CMB17a}$ & 001100+ \\ 
35 & J1306.7-2148 & PKS 1304-215 & rdg & 0.126 & 1.11 & --$^{\rm }$ & 111110+ \\ 
36 & J1325.5-4300 & Cen A & RDG & 0.002 & -0.43 & T$^{\rm ZB95}$ & 111111- \\ 
37 & J1331.0+3032 & 3C 286 & css & 0.850 & 4.70 & C$^{\rm LRL83}$ & 111100+ \\ 
38 & J1346.3-6026 & Cen B & rdg & 0.013 & 0.47 & I$^{\rm JLM01}$ & 111110- \\ 
39 & J1356.2-1726 & PKS B1353-171 & agn & 0.075 & 0.39 & --$^{\rm }$ & 001100+ \\ 
40 & J1443.1+5201 & 3C 303 & rdg & 0.141 & 2.13 & II$^{\rm LRL83}$ & 001110+ \\ 
41 & J1449.5+2746 & B2 1447+27 & rdg & 0.031 & -0.88 & --$^{\rm }$ & 000110+ \\ 
42 & J1449.7-0910 & 1RXS J1449-0910$^e$ & agn & 0.000 & -- & --$^{\rm }$ & 000110+ \\ 
43 & J1459.0+7140 & 3C 309.1 & css & 0.910 & 3.14 & C$^{\rm LRL83}$ & 111100+ \\ 
44 & J1516.5+0015 & PKS 1514+00 & rdg & 0.052 & 0.70 & II$^{\rm CMB17b}$ & 111100+ \\ 
45 & J1518.6+0614 & TXS 1516+064 & rdg & 0.102 & 1.12 & I$^{\rm CMB17a}$ & 000111+ \\ 
46 & J1521.1+0421 & PKS B1518+045 & rdg & 0.052 & 0.42 & I$^{\rm CMB17a}$ & 001110+ \\ 
47 & J1543.6+0452 & CGCG 050-083 & agn & 0.040 & -0.43 & --$^{\rm }$ & 111111+ \\ 
48 & J1630.6+8234 & NGC 6251 & rdg & 0.024 & 0.02 & I/II$^{\rm OL89}$ & 111110+ \\ 
49 & J1724.2-6501 & NGC 6328 & rdg & 0.014 & -0.58 & GPS/CSO$^{\rm T97}$ & 111100- \\ 
50 & J1824.7-3243 & PKS 1821-327 & agn & 0.355 & 3.41 & --$^{\rm }$ & 011100- \\ 
51 & J1829.5+4845 & 3C 380 & css & 0.695 & 4.50 & C$^{\rm LRL83}$ & 111110+ \\ 
52 & J1843.4-4835 & PKS 1839-48 & rdg & 0.111 & 2.21 & I$^{\rm ZB95}$ & 001010- \\ 
53 & J2114.8+2026 & TXS 2112+202 & agn & 0.000 & -- & --$^{\rm }$ & 001110- \\ 
54 & J2156.0-6942 & PKS 2153-69 & rdg & 0.028 & 1.75 & II$^{\rm F98}$ & 011000+ \\ 
55 & J2227.9-3031 & PKS 2225-308 & rdg & 0.056 & -0.08 & I?$^{\rm ZB95}$ & 001010+ \\ 
56 & J2302.8-1841 & PKS 2300-18 & rdg & 0.129 & 1.58 & II?$^{\rm ZB95}$ & 001110+ \\ 
57 & J2326.9-0201 & PKS 2324-02 & rdg & 0.188 & 1.16 & --$^{\rm }$ & 111010+ \\ 
58 & J2329.7-2118 & PKS 2327-215 & rdg & 0.031 & 0.15 & --$^{\rm }$ & 111110+ \\ 
59 & J2334.9-2346 & PKS 2331-240 & agn & 0.048 & 0.63 & II$^{\rm B16}$ & 111100+ \\ 
60 & J2338.1+0325 & PKS 2335+03 & agn & 0.270 & 2.50 & II$^{\rm H17}$ & 011110+ \\ 
61 & J2341.8-2917 & PKS 2338-295 & rdg & 0.052 & -0.58 & --$^{\rm }$ & 001100+ \\ 
 
\enddata
\label{sample1}
\tablecomments{
$a$: Source class in 4FGL-DR2. Capital letters represent objects identified as the same objects found in other wavelengths, and lower case latters represent objects positionally associated with objects found in other wavelengths. \\
$b$: Radio luminosity in units of $10^{24}$ W/Hz at 1.4 GHz from \citet{Angi20} and NED. ``0'' represents that no data is available.\\
$c$: Radio morphology type. I: FR-I, II: R-II, C: core, J: jet. T: transition, H/T: head/tail. Superscript represents references for morphology: 
B16 ; \cite{Bass16}, 
B18 : \cite{Bass18}, 
CMB17a : \cite{Cape17a}, 
CMB17b : \cite{Cape17b}, 
CB88I : \cite{Cond88}, 
D21 : \cite{Das21}, 
F98 : \cite{Fosb98}, 
G94 : \cite{Giov94}, 
H17 : \cite{Hern17}, 
ICS99 : \cite{Ishw99}, 
J82I : \cite{Jenk82}, 
JLM01 : \cite{Jone02}, 
K18 : \cite{Kuzm18}, 
LRL83 : \cite{Lain83}, 
L01 : \cite{Lara01}, 
LJ02 : \cite{Lloy02},. 
M93 : \cite{Mack93}, 
OL89 : \cite{Owen89}, 
P06 : \cite{Puns06}, 
RBC20 : \cite{Rult20}, 
T97 : \cite{Ting97}, 
W04 : \cite{Will04}, 
WH84I : \cite{Wrob84}, 
X95I : \cite{Xu95},
ZB95 : \cite{Zirb95}. 
For CB88I, J82I, WH84I, X95I, we determined FR-I or FR-II based on the radio image in the referance paper.
 \\
$d$: Detection (1) or non-detection (0) at each of 6 bands (0.1--0.3, 0.3--1.0, 1.0--3.0, 3.0--10.0, 10.0--30.0, 30.0--300.0 GeV) in 4FGL-DR2. The right end flag is + for $|b|\geq20^{\circ}$ or 0 for $|b|<20^{\circ}$\\
$e$: The exact catalog name is 1RXS J144942.2-091018.\\
}
\end{deluxetable}
%\end{scriptsize}
%\end{landscape}

%\begin{landscape}
%\begin{scriptsize}
%\startlongtable
\begin{deluxetable}{rccccccc}
\tablecaption{X-ray and Gamma-ray Properties of 4FGL-DR2 radio galaxies}
%\hspace{-1.8cm}
\tablehead{
\colhead{No.$^a$} & \colhead{Data$^b$} & \colhead{$\alpha_G^b$} & \colhead{$\alpha_X^c$} & \colhead{$F_G (L_\gamma)^d$} & \colhead{$F_X (L_X)^e$} & $\nu_{\rm HE}^f$ & $L_{\rm HE}^g$ 
}
\startdata
 1 & xmm & 2.15$\pm$0.13 & 1.97$\pm$0.03 & -11.75(42.42) & -12.26(41.91) & 21.9 & 42.1 \\ 
 & 0202190301 & \\ 
 2 & chandra & 2.19$\pm$0.15 & 1.89$\pm$0.07 & -11.73(44.57) & -11.62(44.68) & 21.4 & 44.7 \\ 
 & 5669 & \\ 
 3 & 0 & 2.89$\pm$0.10 & 0 & -11.35(44.81) & 0.00(0.00) & 19.6 & 45.5 \\ 
 4 & xmm & 2.37$\pm$0.11 & 1.16$\pm$0.02 & -11.45(42.34) & -11.91(41.88) & 21.6 & 42.0 \\ 
 & 0305290201 & \\ 
 5 & swift & 1.89$\pm$0.11 & 1.96$\pm$0.23 & -11.47(42.57) & -11.76(42.28) & 24.9 & 41.8 \\ 
 & 00010149002 & \\ 
 6 & rass & 2.15$\pm$0.17 & 2.0 & -11.82(42.21) & -13.15(40.87) & 25.0 & 41.5 \\ 
 7 & f15 & 1.97$\pm$0.05 & 2.32$\pm$0.04 & -11.05(43.23) & -11.92(42.36) & 24.2 & 42.5 \\ 
 8 & xmm & 2.69$\pm$0.34 & 1.91$\pm$0.02 & -11.39(44.31) & -11.35(44.34) & 21.3 & 44.3 \\ 
 & 0306680301 & \\ 
 9 & xmm & 1.85$\pm$0.15 & 2.44$\pm$0.01 & -11.50(42.41) & -11.39(42.52) & 25.0 & 41.7 \\ 
 & 0151560101 & \\ 
10 & f15 & 2.11$\pm$0.01 & 1.73$\pm$0.03 & -9.48(44.39) & -10.92(42.95) & 22.6 & 43.8 \\ 
11 & xmm & 2.07$\pm$0.06 & 1.69$\pm$0.03 & -11.21(41.68) & -12.22(40.66) & 22.5 & 41.1 \\ 
 & 0502070201 & \\ 
12 & chandra & 1.81$\pm$0.12 & 2.28$\pm$0.07 & -11.61(42.37) & -12.04(41.95) & 25.0 & 41.7 \\ 
 & 857 & \\ 
13 & f15 & 2.74$\pm$0.06 & 1.65$\pm$0.02 & -10.84(43.91) & -10.26(44.48) & 20.0 & 44.2 \\ 
14 & f15 & 2.74$\pm$0.04 & 1.75$\pm$0.03 & -10.84(43.56) & -10.33(44.07) & 20.2 & 44.0 \\ 
15 & xmm & 2.43$\pm$0.12 & 1.77$\pm$0.00 & -11.38(43.07) & -10.77(43.68) & 20.4 & 43.3 \\ 
 & 0206390101 & \\ 
16 & chandra & 2.23$\pm$0.14 & 1.51$\pm$0.14 & -11.55(45.88) & -11.49(45.94) & 21.5 & 45.8 \\ 
 & 14996 & \\ 
17 & xmm & 2.46$\pm$0.01 & 1.80$\pm$0.00 & -10.26(44.61) & -10.89(43.98) & 21.6 & 44.3 \\ 
 & 0065760201 & \\ 
18 & f15 & 1.92$\pm$0.03 & 2.25$\pm$0.02 & -10.82(44.04) & -11.09(43.77) & 24.5 & 43.3 \\ 
19 & chandra & 1.77$\pm$0.15 & 2.61$\pm$0.36 & -11.85(42.08) & -11.89(42.03) & 25.0 & 41.5 \\ 
 & 5900 & \\ 
20 & xmm & 2.20$\pm$0.16 & 1.98$\pm$0.03 & -11.80(42.83) & -12.42(42.21) & 22.3 & 42.6 \\ 
 & 0602390101 & \\ 
21 & xmm & 2.49$\pm$0.11 & 1.59$\pm$0.01 & -11.54(45.76) & -11.53(45.78) & 21.2 & 45.5 \\ 
 & 0147670301 & \\ 
22 & chandra & 2.58$\pm$0.14 & 1.37$\pm$0.02 & -11.62(46.15) & -11.64(46.13) & 21.2 & 45.9 \\ 
 & 434 & \\ 
23 & chandra & 2.53$\pm$0.10 & 1.63$\pm$0.13 & -11.47(45.82) & -11.86(45.43) & 21.7 & 45.4 \\ 
 & 15002 & \\ 
24 & chandra & 2.28$\pm$0.06 & 3.77$\pm$0.52 & -11.27(42.80) & -13.18(40.89) & 22.8 & 42.1 \\ 
 & 18038 & \\ 
25 & chandra & 2.12$\pm$0.16 & 2.56$\pm$0.20 & -11.84(41.37) & -12.19(41.03) & 23.0 & 40.8 \\ 
 & 5902 & \\ 
26 & swift & 1.80$\pm$0.09 & 2.11$\pm$0.24 & -11.47(45.19) & -11.53(45.13) & 25.0 & 44.5 \\ 
 & 00085352004 & \\ 
27 & xmm & 1.39$\pm$0.26 & 3.03$\pm$0.18 & -12.29(42.43) & -13.31(41.40) & 23.9 & 42.5 \\ 
 & 0550270101 & \\ 
28 & swift & 2.64$\pm$0.07 & 1.80$\pm$0.35 & -11.17(0.00) & -11.83(0.00) & 0.0 & 0.0 \\ 
 & 00083516001 & \\ 
29 & xmm & 2.00$\pm$0.09 & 2.26$\pm$0.02 & -11.46(42.57) & -11.46(42.57) & 24.3 & 41.8 \\ 
 & 0602200301 & \\ 
30 & chandra & 2.18$\pm$0.12 & 0.02$\pm$0.10 & -11.61(41.82) & -12.54(40.89) & 23.6 & 41.0 \\ 
 & 10389 & \\ 
31 & xmm & 2.08$\pm$0.15 & 0.34$\pm$0.05 & -11.69(41.39) & -12.11(40.97) & 22.1 & 41.1 \\ 
 & 0502120101 & \\ 
32 & chandra & 2.06$\pm$0.03 & 2.19$\pm$0.03 & -10.75(41.85) & -11.74(40.87) & 23.7 & 41.1 \\ 
 & 18838 & \\ 
33 & rass & 2.33$\pm$0.12 & 2.0 & -11.51(42.60) & -12.40(41.71) & 23.0 & 41.8 \\ 
34 & 0 & 1.98$\pm$0.17 & 0 & -11.81(43.45) & 0.00(0.00) & 24.5 & 42.7 \\ 
35 & swift & 2.18$\pm$0.08 & 1.37$\pm$0.67 & -11.27(44.35) & -12.25(43.37) & 23.5 & 43.6 \\ 
 & 00032810001 & \\ 
36 & f15 & 2.64$\pm$0.02 & 1.73$\pm$0.03 & -10.20(41.06) & -8.21(43.05) & 19.7 & 41.5 \\ 
37 & chandra & 2.39$\pm$0.12 & 2.15$\pm$0.17 & -11.66(45.89) & -12.30(45.24) & 22.7 & 45.2 \\ 
 & 15006 & \\ 
38 & xmm & 2.40$\pm$0.04 & 1.14$\pm$0.01 & -10.61(42.96) & -11.12(42.46) & 22.3 & 42.5 \\ 
 & 0092140101 & \\ 
39 & 0 & 2.02$\pm$0.14 & 0 & -11.97(43.17) & 0.00(0.00) & 20.3 & 44.6 \\ 
40 & nustar & 2.05$\pm$0.12 & 1.77$\pm$0.05 & -11.77(43.96) & -11.48(44.25) & 21.1 & 44.2 \\ 
 & 60463048002 & \\ 
41 & swift & 1.54$\pm$0.17 & 1.70$\pm$0.32 & -12.08(42.25) & -12.37(41.97) & 22.4 & 42.3 \\ 
 & 00040619002 & \\ 
42 & swift & 2.07$\pm$0.18 & 2.20$\pm$0.19 & -11.80(0.00) & -12.12(0.00) & 0.0 & 0.0 \\ 
 & 00087620002 & \\ 
43 & chandra & 2.50$\pm$0.06 & 1.51$\pm$0.02 & -11.36(46.25) & -11.53(46.08) & 21.4 & 45.9 \\ 
 & 3105 & \\ 
44 & xmm & 2.54$\pm$0.10 & 1.78$\pm$0.01 & -11.35(43.46) & -11.54(43.27) & 21.1 & 43.3 \\ 
 & 0103860601 & \\ 
45 & xmm & 1.75$\pm$0.17 & 2.13$\pm$0.01 & -11.86(43.57) & -11.03(44.39) & 25.0 & 42.9 \\ 
 & 0018741001 & \\ 
46 & rass & 2.04$\pm$0.15 & 2.0 & -11.83(42.98) & -13.10(41.72) & 23.8 & 42.3 \\ 
47 & rass & 1.87$\pm$0.07 & 2.0 & -11.27(43.31) & -12.22(42.35) & 24.8 & 42.6 \\ 
48 & f15 & 2.37$\pm$0.03 & 1.82$\pm$0.05 & -10.88(43.24) & -11.59(42.53) & 21.8 & 42.8 \\ 
49 & xmm & 2.50$\pm$0.15 & 1.65$\pm$0.02 & -11.61(42.06) & -12.12(41.55) & 21.9 & 41.5 \\ 
 & 0804520301 & \\ 
50 & xmm & 2.28$\pm$0.11 & 1.73$\pm$0.01 & -11.59(45.04) & -11.09(45.54) & 21.0 & 45.3 \\ 
 & 0650591401 & \\ 
51 & swift & 2.42$\pm$0.03 & 1.60$\pm$0.05 & -10.75(46.58) & -11.21(46.12) & 21.8 & 46.1 \\ 
 & 00081221001 & \\ 
52 & chandra & 2.03$\pm$0.16 & 1.41$\pm$0.14 & -11.79(43.71) & -12.87(42.63) & 23.1 & 43.2 \\ 
 & 10321 & \\ 
53 & rass & 2.11$\pm$0.18 & 2.0 & -11.75(0.00) & -12.70(0.00) & 0.0 & 0.0 \\ 
54 & xmm & 2.87$\pm$0.11 & 1.83$\pm$0.00 & -11.44(42.83) & -10.96(43.30) & 20.2 & 43.1 \\ 
 & 0152670101 & \\ 
55 & chandra & 1.98$\pm$0.18 & 0 & -11.91(42.96) & 0.00(0.00) & 24.3 & 42.3 \\ 
 & 5798 & \\ 
56 & swift & 2.18$\pm$0.12 & 1.59$\pm$0.06 & -11.61(44.03) & -11.05(44.59) & 20.6 & 44.3 \\ 
 & 00031729002 & \\ 
57 & xmm & 2.51$\pm$0.14 & 1.75$\pm$0.06 & -11.46(44.54) & -11.91(44.09) & 21.7 & 44.1 \\ 
 & 0405860101 & \\ 
58 & swift & 2.45$\pm$0.13 & 2.0 & -11.33(43.01) & 0.00(0.00) & 22.6 & 42.3 \\ 
 & 00090972001 & \\ 
59 & xmm & 2.50$\pm$0.09 & 1.74$\pm$0.00 & -11.35(43.38) & -10.79(43.94) & 20.4 & 43.6 \\ 
 & 0760990101 & \\ 
60 & 0 & 2.37$\pm$0.14 & 0 & -11.62(44.73) & 0.00(0.00) & 22.9 & 43.9 \\ 
61 & xmmslew & 2.24$\pm$0.14 & 2.0 & -11.71(43.09) & -12.10(42.71) & 24.1 & 42.3 \\ 

\enddata
\label{sample2}
\tablecomments{
$a$: Object Number in table \ref{sample1}.\\
$b$: Satellites for X-ray data analyzed in this paper. ``f15'', ``rass'', or ``xmmslwe'' represents that \citet{Fuka15}, ROSAT ALl Sky Survey Catalog, or XMM-Newton Slew Survey Catalog is referred to. The number in the 2nd line is an observation ID.\\
$c$: Gamma-ray photon index from 4FGL-DR2.\\
$d$: X-ray photon index.\\
$e$: Logarithmic gamma-ray flux in units of erg s$^{-1}$ cm$^{-2}$ in 4FGL-DR2. Value in the parenthesis a logarithmic gamma-ray luminosity in units of erg s$^{-1}$ in 0.1--100 GeV from 4FGL-DR2.\\
$f$: Logarithmic X-ray flux in units of erg s$^{-1}$ cm$^{-2}$ in 2--10 keV. Values in parentheses a logarithmic X-ray luminosity in units of erg s$^{-1}$.\\
$g$: Logarithmic peak frequency in the SED from X-ray to gamma-ray band in units of Hz.\\
$h$: Logarithmic peak luminosity in the SED from X-ray to gamma-ray band in units of erg s$^{-1}$.\\
In table, ``0'' represents that no data is available or value cannot be estimated.
}
\end{deluxetable}
%\end{scriptsize}
%\end{landscape}

\begin{landscape}
\begin{table}
\begin{center}
\caption{Best-fit parameters for the LDDE model}
%\hspace{-1.8cm}
\begin{scriptsize}
\begin{tabular}{lccccccccccccc}
\hline
\hline
model & $\log_{10} A^a$ & $L_*^b$ & $\gamma_1$ & $\gamma_2$ & $p_1$ & $p_2$ & $z_c$ & $\alpha$ & $\tau$ & $\delta$ & $\ln{\mathcal{L}_{\rm max}}^c$ & KS$_zL^d$ & KS$_z^e$  \\ 
\hline
{\tt BLLz} & -11.55$\pm$0.21 & 2.00 & 5.00 & 1.40$\pm$0.05 & 3.40 & -13.00 & $10^{-8.25\pm4.11}$ & 0.045 & 0.00 & 0.00 & -3269.6 & 55.7 & 46.3 \\ 
{\tt FSRQz} & -11.57$\pm$0.19 & 2.00 & 5.00 & 1.41$\pm$0.05 & 3.40 & -13.00 & $10^{-6.38\pm4.17}$ & 0.210 & 0.00 & 0.00 & -3269.6 & 63.2 & 57.3 \\ 
{\tt BLLp2} & -12.98$\pm$0.26 & 2.00 & 5.00 & 1.42$\pm$0.06 & 3.40 & 6.09$\pm$7.07 & 1.36 & 0.045 & 0.00 & 0.00 & -3269.3 & 69.9 & 61.9 \\ 
{\tt FSRQp2} & -13.12$\pm$0.25 & 2.00 & 5.00 & 1.62$\pm$0.06 & 3.40 & 9.09$\pm$14.42 & 0.56 & 0.210 & 0.00 & 0.00 & -3270.7 & 47.5 & 44.3 \\ 
{\tt BLL0} & -6.68$\pm$0.09 & -0.13$\pm$0.14 & 1.86 & 0.27 & 7.37 & -2.24 & 1.09 & 0.045 & 0.00 & -4.92 & -3276.0 & 0.1 & 0.2 \\
{\tt FSRQ0} & -4.61$\pm$0.20 & -2.06$\pm$0.17 & 1.58 & 0.21 & 7.35 & -6.51 & 0.56 & 0.210 & 0.00 & 0.00 & -3270.3 & 61.1 & 93.2 \\ 
{\tt BLAZAR0} & -6.34$\pm$0.15 & -0.56$\pm$0.18 & 1.83 & 0.50 & 4.96 & -3.39 & 0.90 & 0.072 & -0.64 & -3.16 & -3269.5 & 72.0 & 74.7 \\
{\tt RLF0} & -10.71$\pm$-0.19 & 2.00 & 5.00 & 1.45$\pm$0.04 & 3.40 & -3.40 & 2.00 & 0.000 & 0.00 & 0.00 & -3290.2 &  0.5 &  0.6 \\ 
\hline
\end{tabular}
\end{scriptsize}
\label{lf1}
\end{center}
$a$: In the units of Mpc$^{-3}$.\\
$b$: In the units of $10^{44}$ erg s$^{-1}$.\\
$c$: Maximum likelihood.
$d$: KS probability in units of percent for luminosity distribution.
$e$: KS probability in units of percent for redshift distribution.
\end{table}
\end{landscape}

\begin{landscape}
\begin{table}
\begin{center}
\caption{Best-fit parameters for the {\tt BLLz} model obtained in each of 6 bands}
%\hspace{-1.8cm}
\begin{tabular}{lrcccccccc}
\hline
\hline
$E_{\rm band}$ & $N_{\rm gal}^a$ & $\log_{10} A^b$ & $\gamma_2$ & $z_c$ & $\mu$ & $\sigma$ & $\beta$ & $\ln{\mathcal{L}_{\rm max}}^c$ \\ 
\hline
0.1--0.3 & 13 & -10.78$\pm$0.26 & 1.33$\pm$0.07 & $10^{-7.6\pm3.4}$ & & &  & -1284.9 \\ 
0.3--1 & 19 & -11.13$\pm$0.22 & 1.36$\pm$0.07 & $10^{-4.6\pm2.6}$ & & &  & -1855.2 \\ 
1--3 & 34 & -11.55$\pm$0.21 & 1.40$\pm$0.05 & $10^{-10.0\pm4.1}$ & & &  & -3269.6 \\ 
3--10 & 33 & -11.93$\pm$0.23 & 1.47$\pm$0.06 & $10^{-4.9\pm2.4}$ & & &  & -3149.7 \\ 
10--30 & 27 & -12.00$\pm$0.28 & 1.50$\pm$0.07 & $10^{-1.6\pm4.5}$ & & &  & -2571.0 \\ 
30--300 & 8 & -11.87$\pm$-0.42 & 1.48$\pm$0.10 & $10^{-1.8\pm4.8}$ & & &  & -772.9 \\ 
\hline 
 0.1--300 & 34 & -11.50$\pm$0.11 & 1.42$\pm$0.03 & $10^{-9.7\pm-1.9}$ & 2.26$\pm$0.02 & 0.25$\pm$0.01 & & -17765.1 \\ 
0.1--300 & 34 & -11.62$\pm$0.23 & 1.52$\pm$-0.04 & $10^{-0.3\pm0.3}$ & 2.30$\pm$0.02 & 0.24$\pm$0.01 & 0.06$\pm$0.01 & -17731.7 \\

\hline
\end{tabular}
\label{lf2}
\end{center}
$a$: The number of galaxies used in the fitting.\\
$b$: In units of Mpc$^{-3}$.\\
$c$: Maximum likelihood.
\end{table}
\end{landscape}

\end{document}